\magnification=\magstep1


\def\pd{probability distribution}

\def\ito{It\^o}

\def\real{{\pmb{R}}}

\def\ep{\epsilon}

\def\damtp{\baselineskip=10pt
\eightsl Department of Applied Mathematics and Theoretical Physics,\break%
\eightsl University of Cambridge, Cambridge CB3 9EW, UK\par}
\def\ont{\baselineskip=10pt
\eightsl Optique nonlin\'eaire th\'eorique, \break
Universit\'e libre de Bruxelles, CP231 \break
Bruxelles 1050 BELGIUM}
\def\and{\ifnyasrefstyle \&\ \else and \fi}%
\def\regime{r\'egime}%
%
%

\def\mle{\mu |\ln \epsilon |}
\def\Or#1{{\cal O}(#1)}
\def\mean#1{{\big<#1\big>}}

\def\ha{{1 \over 2}}



\font\sixrm=cmr6           \font\sevenrm=cmr7
\font\sixi=cmmi6           \font\seveni=cmmi7
\font\sixsy=cmsy6          \font\sevensy=cmsy7

\font\eightrm=cmr8         \font\tenrm=cmr10
\font\eighti=cmmi8         
\font\eightsy=cmsy8        
\font\eightit=cmti8        \font\tenit=cmti10
\font\eightsl=cmsl8        \font\tensl=cmsl10
\font\eightbf=cmbx8        \font\tenbf=cmbx10
\font\eighttt=cmtt8        

\font\twelverm=cmr12

\font\twelveit=cmti12
\font\twelvesl=cmsl12
\font\twelvebf=cmbx12

\font\fourteenrm=cmr12 at 14pt

\font\fourteenit=cmti12 at 14pt
\font\fourteensl=cmsl12 at 14pt
\font\fourteenbf=cmbx12 at 14pt




\def\pmb#1{\setbox0=\hbox{#1}%
 \kern-.025em\copy0\kern-\wd0%
 \kern.05em\copy0\kern-\wd0%
 \kern-.025em\raise.0433em\box0}
\def\Bigtype{%
\let\rm=\fourteenrm \let\bf=\fourteenbf%
\let\it=\fourteenit \let\sl=\fourteensl%
\baselineskip=1.44\baselineskip \rm}%
\def\bigtype{%
\let\rm=\twelverm \let\bf=\twelvebf%
\let\it=\twelveit \let\sl=\twelvesl%
\baselineskip=1.2\baselineskip \rm}%
\def\medtype{%
\let\rm=\tenrm \let\bf=\tenbf%
\let\it=\tenit \let\sl=\tensl%
\baselineskip=12pt \rm}%
\def\smalltype{%
\let\rm=\eightrm \let\bf=\eightbf%
\let\it=\eightit \let\sl=\eightsl%
\let\tt=\eighttt
\def\cal{\fam2 }%
 \textfont0=\eightrm   \scriptfont0=\sevenrm \scriptscriptfont0=\sixrm
 \textfont1=\eighti    \scriptfont1=\seveni  \scriptscriptfont1=\sixi
 \textfont2=\eightsy   \scriptfont2=\sevensy \scriptscriptfont2=\sixsy
 \textfont3=\tenex     \scriptfont3=\tenex   \scriptscriptfont3=\tenex
 \textfont\itfam=\eightit
 \textfont\slfam=\eightsl
 \textfont\bffam=\eightbf
 \textfont\ttfam=\eighttt
\baselineskip=10pt
\parskip=1pt plus1pt minus 1pt%
\rm}%
\voffset=0truein%
\vsize=9.3truein%
\hoffset=0.1truein%
\hsize=6.0truein%
\parindent=0.375truein%
\widowpenalty=300

\def\spaceandathird{%
\baselineskip=14pt
\parskip=2pt plus2pt minus2pt%
\everydisplay={\openup 1\jot}}%

\def\singlespace{%
\baselineskip=12pt
\parskip=1pt plus1pt minus 1pt%
\everydisplay={\openup 0\jot}}%

\def\closeup{\singlespace}%

\newdimen\pagewidth
\newdimen\pageheight    
\pagewidth=6.0truein
\vsize=9.3truein%
\pageheight=9.3truein
\newif\ifwindowcover%
\windowcoverfalse%
\outer\def\windowcover{\windowcovertrue}%
\newif\ifcam%
\camfalse%
\outer\def\damtpaddress{\camtrue}%
\newif\ifnofigs%
\nofigsfalse%
\newif\ifchapstart
\chapstartfalse
\def\chapstartpage{\global\chapstarttrue}
\newif\ifinappendices%
\inappendicesfalse%
\newif\iffiguresatend%
\newif\ifnumbereqnsseq%
\def\numbereqnsseq{ \numbereqnsseqtrue}%
\newif\ifnumberfigsseq%
\def\numberfigsseq{ \numberfigsseqtrue}%
\newif\ifnumbertablesseq%
\def\numbertablesseq{ \numbertablesseqtrue}%
\newif\ifnumberrefsintext%
\def\numberrefsintext{ \numberrefsintexttrue}%
\newif\ifinrefs%
\inrefsfalse%
\newif\ifpaperstyle%
\outer\def\paperstyle{\paperstyletrue
     \refauthorinitialssurname%
     \numbereqnsseq \numberfigsseq \numbertablesseq%
     \numberrefsintext}%
\newif\ifsendstyle%
\newif\ifnyasrefstyle%

\newif\ifsinglesided%
\newif\ifdated%

\newif\ifsastyle%
\outer\def\sastyle{\sastyletrue\numberrefsintext%
\ifsinglesided\headline={\rightheadline}
\else\headline={\ifodd\pageno\rightheadline
                        \else\leftheadline\fi}\fi%
\footline={\ifdated\hfil\eightrm\today\else{}\fi}}
\newif\ifheadset%
\headsetfalse%

\countdef\sectno=1%
\countdef\subsectno=2%
\countdef\subsubsectno=3%
\newcount \equano%
\newcount \figno%
\newcount \tableno%
\newcount \referencecount%
\newbox\TitlePage%
\newbox\DedicationPage%
\newbox\PrefacePage%
\newbox\ConclusionPage%
\newbox\AcknowledgementsPage%
\newbox\SummaryPage%
\newbox\TableOfContents%
\newbox\ListOfFigures%
\newbox\Nomenclature%
\newbox\ListOfTables%
\newbox\FiguresPage%
\newbox\figurebox%
\def\thesectno{%
\ifinappendices%
  \ifcase\sectno\or%
    A\or B\or C\or D\or E\or F\or G\or H\or I\or J\or K\or L\or M%
    N\or O\or P\or Q\or R\or S\or T\or U\or V\or W\or X\or Y\or Z\fi%
\else%
  \the\sectno%
\fi}%
\def \lastsubsectno{\thesectno .\the\subsectno}%
\def \lastsubsubsectno{\thesectno .\the\subsectno .\the\subsubsectno}%
\def \lasteqno{\ifnumbereqnsseq (\the\equano)%
\else (\thesectno.\the\equano)\fi}%
\def \preveqno{\advance \equano by -1 \lasteqno}%
\def \lastfigno{\ifnumberfigsseq \hbox{Figure}~\hbox{\the\figno}%
\else \hbox{Figure}~\hbox{\thesectno -\the\figno}\fi}
\def \lasttableno{\ifnumbertablesseq \hbox{Table}~\hbox{\the\tableno}%
\else \hbox{Table}~\hbox{\thesectno .\the\tableno}\fi}%
\def \nextsubsectno{{\advance \subsectno by 1 \lastsubsectno}}
\def \nextsubsubsectno{{\advance \subsubsectno by 1 \lastsubsubsectno}}%
\def \nexteqno{{\advance \equano by 1 \lasteqno}}%
\def \nextfigno{{\advance \figno by 1 \lastfigno}}%
\def \nextnextfigno{{\advance \figno by 2 \lastfigno}}%
\def \nexttableno{{\advance \tableno by 1 \lasttableno}}%
\def\minimalsectno{\ifnum\subsubsectno>0 \lastsubsubsectno%
\else\ifnum\subsectno>0 \lastsubsectno%
\else \thesectno \fi\fi}%
\def\beginsect#1\par{\begingroup
\ifsastyle\thesisheading{#1}
        \chapstartpage
    \xdef\rhead{\ #1}
\else%
    \global \advance \sectno by 1%
    \mainheading{#1}%
\fi
\global \subsectno=0 \global \subsubsectno=0%
\ifnumbereqnsseq \else \global \equano=0 \fi%
\ifnumberfigsseq \else \global \figno=0 \fi%
\ifnumbertablesseq \else \global \tableno=0 \fi%
\message{\thesectno . #1}%
\nobreak\endgroup%
\ifsastyle\headsetfalse\else\fi}%

\def\saheading#1{\vfill\supereject
        \global \advance \sectno by 1%
        \chapstartpage
        \vskip 0pt plus.3\vsize \penalty-100 \vskip 0pt plus-.3\vsize%
        \ifinrefs\noindent{\bigtype \bf #1}
        \else\ifinappendices\noindent{\bigtype \bf #1}
        \else\noindent{\bigtype \the\sectno .\ \bf #1}\fi\fi
        \par\nobreak\noindent}

\def\thesisheading#1{\vfill\supereject
        \global \advance \sectno by 1%
        \vskip 0pt plus.3\vsize \penalty-100 \vskip 0pt plus-.3\vsize%
        \ifinrefs\noindent{\bigtype \bf #1}
        \else\ifinappendices\noindent{\bigtype \bf Appendix \thesectno .\ #1}
        \else\line{}\vskip 0.5truein
        \centerline{\twelverm CHAPTER \the\sectno}
                \vskip 0.5truein
        \centerline{\fourteenbf #1}
        \vskip 0.5truein
        \fi\fi
        \par\nobreak\noindent}

\def\mainheading#1{\begingroup\nohyphens\hbadness=3000%
        \ifinrefs%
          \noindent {\bf #1}\nobreak%
        \else%
           \ifinappendices%
              \noindent {\bf #1}\par\nobreak%
           \else
              \noindent {\bf \thesectno . #1}\par\nobreak%
           \fi%
        \fi%
    \noindent\endgroup}%

\def\subheading#1{\begingroup\nohyphens\hbadness=3000%
\bigskip
    \noindent {\sl\bf\lastsubsectno . #1}\par\nobreak\noindent%
\endgroup}%

\def\subsubheading#1{\begingroup\nohyphens\hbadness=3000%
  \vskip 0pt plus.1\vsize \penalty-100 \vskip 0pt plus-.1\vsize%
\medskip\vskip\parskip%
  \noindent {\sl\lastsubsubsectno . #1}\par\nobreak\smallskip\noindent%
\endgroup}%

\outer \def \beginappendix#1\par{%
\inrefsfalse%
\ifinappendices\else\inappendicestrue\global\sectno=0\fi%
\begingroup%
\ifsastyle\saheading{#1}%
\xdef\rhead{#1}
\else%
    \global \advance \sectno by 1%
    \mainheading{#1}%
    \ifnumbereqnsseq \else \global \equano=0 \fi%
    \ifnumberfigsseq \else \global \figno=0 \fi%
    \ifnumbertablesseq \else \global \tableno=0 \fi%
    \global \subsectno=0 \global \subsubsectno=0%
\fi%
\message{\thesectno . #1}%
\endgroup%
\ifsastyle\headsetfalse\else\fi%
}

\outer \def \beginsubsect#1\par{
\global \advance \subsectno by 1%
\global \subsubsectno=0%
\accretetoTOCtwo{\lastsubsectno.}{#1}%
\subheading{#1}}

\outer \def \beginsubsubsect#1\par{\global \advance \subsubsectno by 1%
\accretetoTOCthree{\lastsubsubsectno.}{#1}\subsubheading{#1}}%
\def \myeqno{\global \advance \equano by 1 {\rm \lasteqno}}%
\def \myeqlabel#1{\expandafter\xdef\csname #1\endcsname{\lasteqno}}

\def \mytextlabel#1{\expandafter\xdef\csname #1\endcsname{\minimalsectno}}
\def \myChapterlabel#1#2{%
 \expandafter\xdef\csname #1\endcsname{Chapter~#2}}
\def \myAppendixlabel#1#2{%
 \expandafter\xdef\csname #1\endcsname{Appendix~\hbox{#2}}}
\def \myfigurelabel#1{\expandafter\xdef\csname #1\endcsname{\lastfigno}}
\def \mytablelabel#1{\expandafter\xdef\csname #1\endcsname{\lasttableno}}
\input epsf

\def\figure #1 #2 #3\par{\global \advance \figno by 1%
\iffiguresatend%
   \global\setbox\ListOfFigures=\vbox{\unvbox\ListOfFigures%
   \bigskip\caption{\hbox{\bf\lastfigno. }{#3}}
   \par\filbreak}%
   \ifnofigs%
   \else
      \global\setbox\FiguresPage=\vbox{\unvbox\FiguresPage%
        \epsfysize=#1
        \centerline{\epsffile{#2}}
       \vskip 0.5truein\relax%
       \hbox{\bf \lastfigno}
        {#3}
        \vfill\eject}
   \fi%
\else%
\ifsastyle
   \global\setbox\ListOfFigures=\vbox{\unvbox\ListOfFigures%
   \bigskip\caption{\hbox%
{\bf\lastfigno. }
{page \folio},
{size in text #1},
{file #2}\filbreak\noindent{#3}}
   \par\filbreak}%
\fi
  \setbox\figurebox=\vbox{%
  \ifnofigs
  \else
        \epsfysize=#1   
        \centerline{\epsffile{#2}}
  \fi
    \caption{{\bf\lastfigno. }{\eightsl #3}}}
  \topinsert\unvbox\figurebox\endinsert%
\fi}%

\def\twofigures #1 #2 #3 #4\par{\global \advance \figno by 1%
\iffiguresatend 
   \begingroup%
   \global\setbox\ListOfFigures=\vbox{\unvbox\ListOfFigures%
   \bigskip\caption{\hbox{\bf\lastfigno. }{#4}}
   \par\filbreak}%
   \ifnofigs
   \else
      \global\setbox\FiguresPage=\vbox{\unvbox\FiguresPage%
        \epsfysize=#1
        \centerline{\epsffile{#2}\hfil\epsffile{#3}}
       \vskip 0.5truein\relax%
       \hbox{\bf \lastfigno}
        \vfill\eject}
   \fi
   \endgroup%
\else%
  \setbox\figurebox=\vbox{%
  \ifnofigs
  \else
        \centerline{
        \epsfysize=#1   
        (a)\epsffile{#2}
        \ \ 
        \epsfysize=#1   
        (b)\epsffile{#3}}
  \fi
    \caption{{\bf\lastfigno. }{\eightsl #4}}}
  \topinsert\unvbox\figurebox\endinsert%
\fi}%

\def\fourfigures #1 #2 #3 #4 #5 #6\par{\global \advance \figno by 1%
\iffiguresatend 
   \begingroup%
   \global\setbox\ListOfFigures=\vbox{\unvbox\ListOfFigures%
   \bigskip\caption{\hbox{\bf\lastfigno. }{#6}}
   \par\filbreak}%
   \ifnofigs
   \else
      \global\setbox\FiguresPage=\vbox{\unvbox\FiguresPage%
        \centerline{
        \epsfysize=#1   
        (a)\epsffile{#2}
        \ \ 
        \epsfysize=#1   
        (b)\epsffile{#3}}
        \line{ }
        \centerline{
        \epsfysize=#1   
        (c)\epsffile{#4}
        \ \ 
        \epsfysize=#1   
        (d)\epsffile{#5}}
        \hbox{\bf \lastfigno}
         {#6}
        \vfill\eject}
   \fi
   \endgroup%
\else%
  \setbox\figurebox=\vbox{%
  \ifnofigs
  \else
        \centerline{
        \epsfysize=#1   
        (a)\epsffile{#2}
        \ \ 
        \epsfysize=#1   
        (b)\epsffile{#3}}
        \line{ }
        \centerline{
        \epsfysize=#1   
        (c)\epsffile{#4}
        \ \ 
        \epsfysize=#1   
        (d)\epsffile{#5}}
  \fi
    \caption{{\bf\lastfigno. }{\eightsl #6}}}
  \topinsert\unvbox\figurebox\endinsert%
\fi}%

\def\eightfigures #1 #2 #3 #4 #5 #6 #7 #8 #9%
\par{\global \advance \figno by 1%
\iffiguresatend 
   \begingroup%
   \global\setbox\ListOfFigures=\vbox{\unvbox\ListOfFigures%
   \bigskip\caption{\hbox{\bf\lastfigno. }{#6}}
   \par\filbreak}%
   \ifnofigs
   \else
      \global\setbox\FiguresPage=\vbox{\unvbox\FiguresPage%
        \epsfysize=#1
        \centerline{\epsffile{#2}\hfil\epsffile{#3}}
       \vskip 0.5truein\relax%
       \hbox{\bf \lastfigno}
        \vfill\eject}
   \fi
   \endgroup%
\else%
  \setbox\figurebox=\vbox{%
  \ifnofigs
  \else
        \centerline{
        \epsfysize=1.8truein
        (a)\epsffile{#1}
        \epsfysize=1.8truein
        (b)\epsffile{#2}
        \epsfysize=1.8truein
        (c)\epsffile{#3}}
        \line{ }
        \centerline{
        \epsfysize=1.8truein
        (d)\epsffile{#4}
        \epsfysize=1.8truein
        (e)\epsffile{#5}
        \epsfysize=1.8truein
        (f)\epsffile{#6}}
        \line{ }
        \centerline{
        \epsfysize=1.8truein
        (g)\epsffile{#7}
        \epsfysize=1.8truein
        (h)\epsffile{#8}
        \epsfysize=1.8truein
        \ \ \ \epsffile{key.eps}}
  \fi
    \caption{{\bf\lastfigno. }{\eightsl #9}}
}
  \topinsert\unvbox\figurebox\endinsert%
\fi}%

\def\caption#1{%
\smallskip%
\begingroup
\nohyphens%
\advance \leftskip by \parindent%
\advance \rightskip by \parindent
\spaceskip=.3333em plus .3333em \xspaceskip=.5em plus .5em%
\par \noindent{%
\iffiguresatend{#1}\else\smalltype{#1}\par\fi\par} \par\endgroup}%

\outer \def \beginrefs{\inrefstrue\inappendicesfalse}%

\def\showrefs{\ifsastyle\accretetoTOCone{}{References}\headsetfalse\else\fi%
\setbox\myrefs=\vbox{\unvbox\myrefs\vfil}%
\ifsendstyle\vfill\eject\else\smallskip\goodbreak\fi
    \unvbox\myrefs}

\def\addtorefs#1#2{%
\global\advance\referencecount by 1%
\expandafter\xdef\csname#2\endcsname{\the\referencecount}%
\global\setbox\myrefs=\vbox{\unvbox\myrefs%
{\ifsendstyle\else\closeup\fi
\ifnyasrefstyle
\item{\csname#2\endcsname.}\frenchspacing{\csname#1\endcsname}
\else
\item{[\csname#2\endcsname]}\frenchspacing{\csname#1\endcsname}
\fi
\hfil\par\ifsendstyle\medskip\fi}}}

\def\addref#1#2{\ifundefined{#2}\addtorefs{#1}{#2}\fi
\ifnyasrefstyle$^{\csname#2\endcsname}$\else~[\csname#2\endcsname]\fi}%


\def\addtworefs#1#2#3#4{%
{\ifundefined{#2}\addtorefs{#1}{#2}\fi}%
{\ifundefined{#4}\addtorefs{#3}{#4}\fi}%
\ifnyasrefstyle
$^{\csname#2\endcsname,~\csname#4\endcsname}$%
\else~[\csname#2\endcsname,~\csname#4\endcsname]\fi}%

\def\addthreerefs#1#2#3#4#5#6{%
{\ifundefined{#2}\addtorefs{#1}{#2}\fi}%
{\ifundefined{#4}\addtorefs{#3}{#4}\fi}%
{\ifundefined{#6}\addtorefs{#5}{#6}\fi}%
\ifnyasrefstyle
$^{\csname#2\endcsname,\csname#4\endcsname,\csname#6\endcsname}$%
\else~[\csname#2\endcsname,\csname#4\endcsname,\csname#6\endcsname]\fi}%

\def\addfourrefs#1#2#3#4#5#6#7#8{
{\ifundefined{#2}\addtorefs{#1}{#2}\fi}%
{\ifundefined{#4}\addtorefs{#3}{#4}\fi}%
{\ifundefined{#6}\addtorefs{#5}{#6}\fi}%
{\ifundefined{#8}\addtorefs{#7}{#8}\fi}%
\ifnyasrefstyle
$^{\csname#2\endcsname-\csname#8\endcsname}$%
\else~[\csname#2\endcsname-\csname#8\endcsname]\fi}%


\def\genericref#1#2{\begingroup\closeup%
\expandafter\xdef\csname#1\endcsname{#2.}%
\endgroup}%
\def\refyear#1{\ifnyasrefstyle #1. \else (#1)\fi}%
\def\reftitle#1{\lq\lq#1''}%
\def\refauthorinitialssurname{%
\def\author##1##2{##1~##2}
\def\authorn##1##2{\ifnyasrefstyle\uppercase{##2,~##1}\else{##1~##2}\fi}}%
\refauthorinitialssurname%
\outer\def\refarticle#1#2#3#4#5#6#7{%
\genericref{#1}{%
\ifnyasrefstyle
    {#2.\ \refyear{#3} \reftitle{#4}. {#5} {\bf #6}: {#7}}%
\else%
  \ifpaperstyle%
    {#2\ \reftitle{#4.} {\sl #5} {\bf #6} #7\ \refyear{#3}}%
  \else
    {#2\ \reftitle{#4.} {\sl #5} {\bf #6} #7\ \refyear{#3}}%
  \fi%
\fi}}%

\outer\def\refarticletobe#1#2#3#4#5{%
\genericref{#1}{#2, \reftitle{#4} {#5} \refyear{#3}}}

\outer\def\refbook#1#2#3#4#5#6{%
\genericref{#1}%
{\ifnyasrefstyle%
#2.\ #3. #4. #5, #6%
\else%
\ifpaperstyle%
{#2,\ {\sl #4}\ (#5, \ #6, \ #3)}%
  \else%
    {#2\ {\sl #4}.  #5: #6 \ \refyear{#3}}%
\fi\fi}}%

\outer\def\refartbook#1#2#3#4#5#6#7#8#9{%
\genericref{#1}%
{\ifpaperstyle%
    {#2 \refyear{#3} \reftitle{#4,} in {\sl #5}, #6 (Ed.). #7: #8}%
\else%
    {#2 \reftitle{#4} #9 in {\sl #5}, Ed. #6. #7: #8 \refyear{#3}}%
\fi}}%

\outer\def\refphd#1#2#3#4#5{%
\genericref{#1}%
{\ifpaperstyle%
  {#2, {\sl #4} (Ph.D. thesis, #5, #3)}%
\else%
    {#2 \refyear{#3} \reftitle{#4.} {#5, Ph.D. thesis}}%
\fi}}

\outer\def\refnote#1{%
\genericref{#1}}
\def\theauthor{Grant Lythe}

\def\title#1\par{\message{#1}%
\xdef\lhead{#1}%
\xdef\rhead{}%
\begingroup
  \ifwindowcover%
    \dimen1=2.5truein \advance\dimen1 by -\voffset%
    \global\setbox\TitlePage=\vbox{\vbox to \dimen1{%
      \dimen2=1.75truein \advance\dimen2 by -\voffset%
      \smallskip \relax%
      \dimen3=1.5truein \advance\dimen3 by -\hoffset%
      \leftskip=\dimen3 plus3em\relax%
      \dimen4=\hsize \advance\dimen4 by \hoffset \advance\dimen4 by -5truein%
      \rightskip=\dimen4 plus3em\relax%
      \parfillskip=0pt \parindent=0pt \spaceskip=0.3333em
        \xspaceskip=0.5em%
        \line{}\bigskip\line{}\bigskip
      \centerline{\bigtype\bf #1}\smallskip
          {\centerline{\theauthor}}
        }}
  \fi%
\endgroup}%

\def\signaturepaper{
\global\setbox\TitlePage=\vbox{%
\unvbox\TitlePage%
\noindent\narrower\raggedcenter{\bigtype%
\ifwindowcover \else \theauthor\par\fi%
\ifcam\damtp\else\ont\fi
\bigskip}}}%

\def\signaturethesis{%
\parskip=20pt plus 10pt%
\global\setbox\TitlePage=\vbox{%
\line{}\bigskip
\vfill%
\noindent\narrower\raggedcenter{%
\bigtype%
{\bf Stochastic slow-fast dynamics}\par%
\bigtype%
Grant David Lythe\par%
\medtype
Trinity College, Cambridge\par%
\vfill%
Dissertation submitted\break%
for the degree of\break%
Doctor of Philosophy\break%
at the\break%
University of Cambridge\break%
\vfill%
October 1994}}}%

\def\signature{\begingroup%
\ifpaperstyle \signaturepaper \fi%
\ifsastyle \signaturethesis \fi%
\endgroup}%

\long\def\addtotitlepage#1{
\global\setbox\TitlePage=\vbox{%
\unvbox\TitlePage%
\bigskip\noindent#1\par}}%

\def\abstract#1\par{\begingroup%
\addtotitlepage{{\centerline{\bf Abstract}}\par\nobreak\medskip\noindent#1}%
\endgroup}%

\def\submittedto#1{\begingroup\closeup%
\addtotitlepage{\smalltype
\ifpaperstyle  {Submitted to  \hfill printed on \ \today}\fi%
}\endgroup}%

\def\publishedin#1{\begingroup\closeup%
\addtotitlepage{\bigtype
\ifpaperstyle  {#1}\fi}\endgroup}%

\long\def\summary#1{%
\global\setbox\SummaryPage=\vbox{%
  \unvbox\SummaryPage \noindent #1\par}%
  \begingroup\pageno=-4\endgroup}

\def\Showbox#1{\begingroup%
\loop%
\ifvbox#1%
    \dimen255=\vsize%
    \advance \dimen255 by -\topskip%
    \setbox0=\vsplit#1 to \dimen255%
    \unvbox0%
    \null \vfill \penalty -10000%
\repeat%
\endgroup}%

\def\ShowTP{\begingroup \nopagenumbers\headline={}%
\ifvbox\TitlePage%
    \global \sectno=0 \global \subsectno=0 \global \subsubsectno=0%
    \setbox\TitlePage=\vbox{\unvbox\TitlePage\vfil}%
    \Showbox\TitlePage%
\fi \endgroup}%

\def\PreambleHeader#1{%
\begingroup\raggedcenter\line{}\vskip 0.3truein 
\ifsastyle\bigtype\else\Bigtype\fi\bf \noindent%
#1\par \vskip 0.2truein\endgroup\noindent}%

\def\ShowAck{%
\ifvbox\AcknowledgementsPage%
    \global \sectno=0 \global \subsectno=0 \global \subsubsectno=0%
    \setbox\AcknowledgementsPage=\vbox{%
\PreambleHeader{Acknowledgements}\unvbox\AcknowledgementsPage \vfil}%
    \Showbox\AcknowledgementsPage%
\fi}%

\def\ShowSummary{%
\ifvbox\SummaryPage%
    \global \sectno=0 \global \subsectno=0 \global \subsubsectno=0%
    \setbox\SummaryPage=\vbox{%
\PreambleHeader{Summary}\unvbox\SummaryPage \vfil}%
    \Showbox\SummaryPage%
\fi}%

\def\Trailers{%
\pageno=-1%
\footline={}
\ifsastyle\headline={}%
 \ShowTOC \fi
\iffiguresatend \begingroup \nopagenumbers \ShowLOF \ShowFP \endgroup\fi%
}%
\def\themonth{\ifcase\month\or%
  January\or February\or March\or April\or%
  May\or June\or July\or August\or%
  September\or October\or November\or December\fi}%
\def\today{\number\day%
\space%
\themonth%
\space%
\number\year}%
%
\outer\def\raggedright{\rightskip=0pt plus 4em%
  \spaceskip=.3333em \xspaceskip=.5em%
  \pretolerance=200 \tolerance=400}%
\def\leaderfill{\leaders\hbox to 1em{\hss.\hss}\hfill}%
\def\dddot#1{\begingroup\vbox{%
\moveleft 0.15em\hbox{...}%
\nointerlineskip\vskip0.4ex%
\hbox{$\displaystyle{#1}$}}\endgroup}%
\def\ifundefined#1{\expandafter\ifx\csname#1\endcsname\relax}%
\hyphenation{co-dimen-sion sol-ution sol-utions sun-spot sun-spots
pro-pen-sity}
\overfullrule=0pt%
\def\nohyphens{\pretolerance=9999 \tolerance=9999%
\hyphenpenalty=9999 \exhyphenpenalty=1000 }%
\def\raggedcenter{\leftskip=0pt plus6em \rightskip=\leftskip%
\parfillskip=0pt \parindent=0pt \spaceskip=0.3333em \xspaceskip=0.5em%
\nohyphens \hbadness=3000}%
%
%
%
%
%
\outer\def \table#1\par{\global \advance \tableno by 1%
\begingroup \closeup
\global\setbox\ListOfTables=\vbox{\unvbox\ListOfTables%
\smallskip\noindent{\bf\lasttableno.\ }{\strut\sl #1\strut}\par}\endgroup%
\medskip\caption{{\bf\lasttableno.\ }{\sl #1}}}%

\paperstyle

\newbox\myrefs%
    \setbox\myrefs=\vbox{%
\ifsastyle\noindent{\bigtype \bf{References}\medskip}
\else\smallskip\noindent\bf References\par\medskip\fi}%

\beginrefs 
\begingroup\raggedright%
\nohyphens%

\refarticle{refA11}
{\author{V.S.}{Afra\u\i movich}, \author{V.V.}{Bykov} \and
 \author{L.P.}{Shil'nikov}}{1983}
{On structurally unstable attracting limit sets of Lorenz attractor type}
            {Trans. Moscow Math. Soc.}{44}{153--216}
 
\refartbook{refA19}
{\author{V.S.}{Afra\u\i movich} \and \author{L.P.}{Shil'nikov}}{1983}
            {Strange attractors and quasiattractors}
            {Nonlinear Dynamics and Turbulence}
            {\author{G.I.}{Barenblatt}, \author{G.}{Iooss} \and
             \author{D.D.}{Joseph}}
            {Pitman}{Boston}{1--34}
 
\refarticle{refA18}
{\author{G.}{Ahlers} \and \author{I.}{Rehberg}}{1986}
            {Convection in a binary mixture heated from below}
            {Phys. Rev. Lett.}{56}{1373--1376}
 
\refarticle{refA21}
{\author{K.H.}{Alfsen} \and \author{J.}{Fr\o yland}}{1985}
            {Systematics of the Lorenz model at $\sigma=10$}
            {Physica Scripta}{31}{15--20}
 
\refarticle{refA22}
{\author{I.S.}{Aranson}, \author{V.S.}{Afra\u\i movich} \and
 \author{M.I.}{Rabinovich}}{1989}
            {Multidimensional strange attractors and turbulence}
            {Sov. Sci. Rev.~C. Math. Phys.}{8}{293--376}
 
\refarticle{refA10}
{\author{A.}{Arn\'eodo}, \author{P.}{Coullet} \and \author{C.}{Tresser}}{1981}
            {A possible new mechanism for the onset of turbulence}
            {Phys. Lett.}{81A}{197--201}
 
\refarticle{refA4}
{\author{A.}{Arn\'eodo}, \author{P.H.}{Coullet} \and \author{E.A.}{Spiegel}}{1982}
            {Chaos in a finite macroscopic system}
            {Phys. Lett.}{92A}{369--373}
 
\refarticle{refA2}
{\author{A.}{Arn\'eodo}, \author{P.H.}{Coullet} \and \author{E.A.}{Spiegel}}
            {1985}
            {The dynamics of triple convection}
            {Geophys. Astrophys. Fluid Dynamics}{31}{1--48}
 
\refarticle{refA3}
{\author{A.}{Arn\'eodo}, \author{P.H.}{Coullet}, \author{E.A.}{Spiegel} \and
 \author{C.}{Tresser}}
            {1985}
            {Asymptotic chaos}
            {Physica}{14D}{327--347}
 
\refarticle{refA13}
{\author{A.}{Arn\'eodo} \and \author{O.}{Thual}}{1985}
            {Direct numerical simulations of a triple convection problem
             versus normal form predictions}
            {Phys. Lett.}{109A}{367--373}
 
\refarticle{refA15}
{\author{V.I.}{Arnol'd}}{1972}
           {Lectures on bifurcations in versal families}
           {Russ. Math. Surveys}{27}{54--123}
 
\refbook{refA14}
{\author{V.I.}{Arnol'd}}{1983}
            {Geometrical Methods in the Theory of Ordinary Differential
             Equations}
            {Springer}{New York}
 
\refarticle{refA9}
{\author{W.}{Arter}}{1983}
            {Nonlinear convection in an imposed horizontal magnetic field}
            {Geophys. Astrophys. Fluid Dynamics}{25}{259--292}
 
\refbook{refA100}
{\author{L.}{Arnold}}{1974}
    {Stochastic Differential Equations}
    {Wiley}{New York}

\refbook{refA101}
{\author{Milton}{Abramowitz} \and \author{Irene}{Stegun}}{1968}
    {Handbook of Mathematical Functions}
    {Dover}{New York}
 
\refarticle{refA102}
{\author{*}{McA**}}{1888}
           {Lognormal stuff}
           {J.~Appl. Math. Phys. (ZAMP)}{37}{608--623}

\refarticle{refA103}
{\author{Guenter}{Ahlers}, \author{Christopher W.}{Meyer}
\& \author{I.}{Rehberg}}{1989}
            {Deterministic and Stochastic Effects Near the Convective Onset}
            {J. Stat. Phys.}{54}{1121--1131}
\refarticle{refA104}
{\author{Guenter}{Ahlers}, \author{M.C.}{Cross}, \author{P.C.}{Hohenberg}
\and \author{S.}{Safran}}{1981}
            {The amplitude equation near the convective threshold:
application to time-dependent heating experiments}
            {J. Fluid Mech.}{110}{297--334}

\refbook{refA105}
{\author{V.I.}{Arnol'd (Ed.)}}{1994}
            {Dynamical Systems V}
            {Springer}{Berlin}
\refbook{refA106}
{\author{Robert J.}{Adler}}{1981}
            {The Geometry of Random Fields}
            {Wiley}{Chichester}
\refbook{refA107}
{\author{Robert J.}{Adler}}{1990}
            {An Introduction to Continuity, Extrema and Related
Topics for General Gaussian Processes}
            {Institute of Mathematical Statistics}{Hayward, California}

\refarticle{refA108}
{\author{Dieter}{Armbruster}, \author{John}{Guckenheimer}
 \and \author{Philip}{Holmes}}{1988}
           {Heteroclinic cycles and modulated travelling waves in
systems with O(2) symmetry}
           {Physica D}{29}{257--282}
 
\refbook{refB2}
{\author{P.}{Berg\'e}, \author{Y.}{Pomeau} \and \author{C.}{Vidal}}{1984}
            {Order Within Chaos}
            {Wiley}{New York}

\refarticle{refB3}
{\author{C.S.}{Bretherton} \and \author{E.A.}{Spiegel}}{1983}
           {Intermittency through modulational instability}
           {Phys. Lett.}{96A}{152--156}
 
\refarticle{refB25}
{\author{F.H.}{Busse} \and \author{A.C.}{Or}}{1986}
           {Subharmonic and asymmetric convection rolls}
           {J.~Appl. Math. Phys. (ZAMP)}{37}{608--623}
 
\refarticle{refB100}
{\author{G.}{Broggi}, \author{A.}{Colombo}, \author{L.A.}{Lugiato}
\and \author{Paul}{Mandel}}{1986}
          {Influence of white noise on delayed bifurcations}
           {Phys. Rev. A}{33}{3635--3637}

\refarticle{refB101}
{\author{C.}{van den Broeck} \and \author{Paul}{Mandel}}{1987}
           {Delayed bifurcations in the presence of noise}
           {Phys. Lett.}{A122}{36--38}

\refarticle{refB102}
{\author{Jean-Marie}{Wersinger}, \author{John M.}{Finn}
  \and \author{Edward}{Ott}}{1980}
            {Bifurcations and strange behaviour in instability
saturation by nonlinear mode coupling}
         {Phys. Rev. Lett.}{44}{453--457}

\refarticle{refB103}
{\author{J.Robert}{Buchler}, \author{Pawe\l}{Moskalik} \and\author{G\'eza}{Kov\'acs}}{1991}
           {Periodic stellar pulsations: stability analysis and
amplitude equations}
           {Ap. J.}{380}{185--199}

\refarticle{refB104}
{\author{J.R.}{Buchler} \and\author{G.}{Kov\'acs}}{1986}
            {The effects of a 2:1 resonance in nonlinear radial
stellar pulsations}
            {Ap. J.}{303}{749--765}

\refartbook{refB105}
{\author{Claude}{Baesens}}{1991}
{Noise effect on dynamic bifurcations: the case of a
           period-doubling cascade}
            {Dynamic Bifurcations}
            {\author{E.}{Beno\^\i t}}
            {Springer}{Berlin}{pp107--130}

\refartbook{refB106}
{\author{E}{Beno\^\i t)}}{1991}
            {Linear dynamic bifurcation with noise}
            {Dynamic Bifurcations}
            {\author{E.}{Beno\^\i t}}
            {Springer}{Berlin}{pp131--150}

\refbook{refB107}
{\author{Denis R.}{Bell}}{1987}
            {The Malliavin Calculus}
            {Longman}{New York}
 
\refarticle{refB108}
{\author{A.R.}{Bulsara} \author{W.C.}{Schieve} \and
\author{E.W.}{Jacobs}}{1990}
            {Homoclinic chaos in a system perturbed by weak Langevin noise}
            {Phys. Rev. A}{41}{668--681}

\refarticle{refB109}
{\author{Per}{Bak}, \author{Chao}{Tang} \and
\author{Kurt}{Wiesenfeld}}{1988}
            {Self-organized criticality}
            {Phys. Rev. A}{38}{364--374}

\refbook{refB110}
{\author{Nicola}{Bellomo} \and \author{Rioccardo}{Riganti}}{1987}
            {Nonlinear Stochastic Systems in Physics and Mathematics}
            {World Scientific}{Singapore}
 
\refarticle{refB111}
{\author{S.M.}{Baer}, \author{T.}{Erneux}}{1986}
            {Singular Hopf bifurcation to relaxation oscillations}
            {SIAM J. Appl. Math.}{46}{721--739}

\refarticle{refB112}
{\author{S.M.}{Baer}, \author{T.}{Erneux}}{1992}
            {Singular Hopf bifurcation to relaxation oscillations II}
            {SIAM J. Appl. Math.}{52}{1651--1664}

\refarticle{refB113}
{\author{Claude}{Baesens}}{1991}
            {Slow sweep through a period-doubling cascade:
Delayed bifurcations and renormalisation}
            {Physica D}{53}{319--375}

\refarticle{refB114}
{\author{Andreas}{Becker} \and \author{Lorenz}{Kramer}}{1994}
            {Linear stability analysis for bifurcations in systems
with fluctuating control parameter}
            {Phys. Rev. Lett.}{73}{955--958}

\refarticle{refB115}
{\author{Paul}{Bryant} \and \author{Kurt}{Wiesenfeld}}{1986}
            {Suppression of period-doubling and nonlinear parametric
effects in periodically perturbed systems}
            {Phys. Rev. A}{33}{2525--2543}

\refbook{refB200}
{\author{E.}{Beno\^\i t (Ed.)}}{1991}
            {Dynamic bifurcations}
            {Springer}{Berlin}
 
\refbook{refC23}
{\author{J.}{Carr}}{1981}
    {Applications of Centre Manifold Theory}
    {Springer}{New York}
 
\refphd{refC30}
{\author{F.}{Cattaneo}}{1984}
            {Compressible magnetoconvection}
            {University Cambridge}
 
\refartbook{refC31}
{\author{F.}{Cattaneo}}{1984}
            {Oscillatory convection in sunspots}
            {The Hydromagnetics of the Sun}
            {\author{T.D.}{Guyenne} \and \author{J.J.}{Hunt}}
            {ESA}{Noordwijkerhoute}{47--50}
 
\refbook{refC22}
{\author{S.}{Chandrasekhar}}{1961}
            {Hydrodynamic and Hydromagnetic Stability}
            {Clarendon Press}{Oxford}
 
\refarticle{refC17}
{\author{J.}{Coste} \and \author{N.}{Peyraud}}{1982}
            {A new type of period-doubling bifurcations in one-dimensional
             transformations with two extrema}
            {Physica}{5D}{415--420}
 
\refarticle{refC3}
{\author{P.H.}{Coullet} \and \author{E.A.}{Spiegel}}{1983}
            {Amplitude equations for systems with competing instabilities}
            {SIAM J.~Appl. Math.}{43}{776--821}
 
\refarticle{refC15}
{\author{P.}{Coullet}, \author{J.-M.}{Gambaudo} \and \author{C.}{Tresser}}{1984}
            {Une nouvelle bifurcation de codimension~2: le collage de cycles}
            {C.~R.~Acad. Sc. Paris}{299}{S\'erie~I, 253--256}
 
\refarticle{refC7}
{\author{J.H.}{Curry}, \author{J.R.}{Herring},
 \author{J.}{Loncaric} \and \author{S.A.}{Orszag}}{1984}
            {Order and disorder in two- and three-dimensional B\'enard
             convection}
            {J.~Fluid Mech.}{147}{1--38}

\refbook{refC100}
{\author{P.}{Collet} \and \author{J.-P.}{Eckmann}}{1986}
    {Iterated maps of the interval as dynamical systems}
    {Birkh\ddot auser}{Boston}
 
\refarticle{refC101}
{\author{J.P.}{Crutchfield}, \author{J.D.}{Farmer}}{1984}
            {Fluctuations and simple chaotic dynamics}
            {Phys. Rep.}{92}{45--82}
\refarticle{refC102}
{\author{Alex}{Craik}}{1992}
            {Second-harmonic resonance in non-conservative systems}
            {Wave Motion}{15}{173--183}

\refphd{refC103}
{\author{Carston C.}{Chow}}{1985}
            {Spatiotemporal chaos on the nonlinear three wave interaction}
            {Massachusetts Institute of Technology}

\refbook{refC104}
{\author{C.}{Canuto et al}}{1988}
    {Spectral Methods in Fluid Dynamics}
    {Springer}{New York}

\refarticle{refC105}
{\author{Hugues}{Chat\'e} and \author{Paul}{Manneville}}{1987}
            {Transition to Turbulence via Spatiotemporal Intermittency}
            {Phys. Rev. Lett.}{58}{112--115}

\refarticle{refC106}
{\author{M.C.}{Cross} \and \author{P.C.}{Hohenberg}}{1993}
            {Pattern formation outside of equilibrium}
            {Rev. Mod. Phys.}{92}{851--1089}

\refarticle{refC107}
{\author{M.}{Ciofini}, \author{R.}{Meucci}
\and \author{F.T.}{Arecchi}}{1990}
            {Transient statistics in a $CO_2$ laser with a slowly
swept pump}
            {Phys. Rev. A}{42}{482--486}

\refarticle{refC108}
{\author{Jack}{Carr} \and \author{Robert}{Pego}}{1992}
            {Self-similarity in a coarsening model in one dimension}
            {Proc. Royal Soc. London A}{436}{569--582}

\refarticle{refC109}
{\author{J.}{Carr} \and \author{R.L.}{Pego}}{1989}
            {Metastable Patterns in Solutions of $u_t=\ep^2u_{xx}-f(u)$}
            {Commun. Pure Appl. Math.}{42}{523--576}

\refarticle{refC111}
{\author{Jack}{Carr} \and \author{Robert}{Pego}}{1992}
            {Invariant manifolds for metastable patterns in
        $u_t=\ep^2u_{xx}-f(u)$}
            {Proc. Royal Soc. Edinburgh A}{116}{133--160}
\refbook{refC110}
{\author{H.S.}{Carslaw} \and \author{J.C.}{Jaeger}}{1959}
    {Conduction of heat in solids}
    {Oxford University Press}{London}

\refarticle{refC111}
{\author{J.P.}{Crutchfield}, \author{B.A.}{Huberman}}{1980}
            {Fluctuations and the onset of chaos}
            {Phys. Lett. A}{77}{407--410}

\refarticle{refC112}
{\author{Thomas}{Clune}, \author{Edgar}{Knobloch}}{1994}
            {pattern selection in three-dimensional magnetoconvection}
            {Physica D}{74}{151--176}

\refarticle{refD2}
{\author{L.N.}{Da~Costa}, \author{E.}{Knobloch} \and \author{N.O.}{Weiss}}{1981}
            {Oscillations in double-diffusive convection}
            {J.~Fluid Mech.}{109}{25--43}
 
\refarticle{refD14}
{\author{G.}{Dangelmayr}, \author{D.}{Armbruster} \and
 \author{M.}{Neveling}}{1985}
            {A codimension three bifurcation for the laser with saturable
             absorber}
            {Z.~Phys.~B -- Condensed Matter}{59}{365--370}
 
\refarticle{refD3}
{\author{G.}{Dangelmayr} \and \author{E.}{Knobloch}}{1986}
            {Interaction between standing and travelling waves and steady
             states in magnetoconvection}
            {Phys. Lett.}{117A}{394--398}
 
\refarticle{refD1}
{\author{G.}{Dangelmayr} \and \author{E.}{Knobloch}}{1987}
            {The Takens--Bogdanov bifurcation with $O(2)$-symmetry}
            {Phil. Trans. R.~Soc. Lond.}{A322}{243--279}
 
\refarticle{refD6}
{\author{A.E.}{Deane}, \author{E.}{Knobloch} \and \author{J.}{Toomre}}{1988}
            {Travelling waves in large-aspect-ratio thermosolutal convection}
            {Phys. Rev.}{37A}{1817--1820}
 
\refarticle{refD11}
{\author{A.E.}{Deane} \and \author{L.}{Sirovich}}{1991}
            {A computational study of Rayleigh--B\'enard convection. Part~1.
             Rayleigh-number scaling}
            {J.~Fluid Mech.}{222}{231--250}
 
\refarticle{refD15}
{\author{P.}{Deuflhard}}{1983}
            {Order and stepsize control in extrapolation methods}
            {Numer. Math.}{41}{399--422}
 
\refbook{refD13}
{\author{E.}{Doedel}}{1986}%
            {AUTO: Software for Continuation and Bifurcation Problems in
             Ordinary Differential Equations}
            {C.I.T. Press}{Pasadena}

\refarticle{refD100}
{\author{M.I.}{Dykman},
\author{R.}{Mannella}, \author{P.V.E.}{McClintock},
\author{N.D.}{Stein} \and \author{N.G.}{Stocks}}{1993}
           {Probability distributions and escape rates for systems
driven by quasimonochromatic noise}
           {Phys. Rev. E}{47}{3996--4009}

\refarticle{refD101}
{\author{I.T.}{Drummond}}{1993}
           {Multiplicative stochastic differential equations with
noise-induced transitions}
           {J. Phys. A}{25}{2273--2296}

\refarticle{refD102}
{\author{S.M.}{Catterall}, \author{I.T.}{Drummond}
\and \author{R.R.}{Horgan}}{1991}
           {Stochastic simulation of quantum mechanics}
           {J. Phys. A}{24}{4081--4091}

\refarticle{refD103}
{\author{Didier}{Dangoisse}, \author{Pierre}{Glorieux}
\and \author{Daniel}{Hennequin}}{1987}
           {Chaos in a $CO_2$ laser with modulated control parameters:
        Experiments and numerical simulations}
           {Phys. Rev. A}{36}{4775--4791}

\refarticle{refD104}
{\author{Charles R.}{Doering}}{1987}
           {Nonlinear Parabolic Stochastic Differential Equations
with Additive Colored Noise on $\real^d\times\real_+$:
A Regulated Stochastic Quantization}
           {Commun. Math. Phys.}{109}{537--561}

\refarticle{refD105}
{\author{Charles R.}{Doering}}{1987}
           {A stochastic partial differential equation
with multiplicative noise}
           {Phys. Lett. A}{122}{133--139}

\refarticle{refD106}
{\author{Marc}{Diener}}{1984}
           {The Canard Unchained or How Fast/Slow Dynamical Systems Bifurcate}
           {The Mathematical Intelligencer}{6}{No.2, 38--49}

\refartbook{refD107}
{\author{C.R.}{Doering}}{1992}
            {Evaluating the Path Integral:
Field Theory Estimates via Stochastic Quantization}
            {Lectures on Path Integration}
            {\author{R.}{Cerdeira}}
            {World Scientific}{Singapore}{pp184--214}

\refartbook{refD108}
{\author{Marc}{Diener}}{1981}
            {On the Perfect Delay Convention, or
the Revolt of the Slaved Variables}
            {Chaos and Order in Nature}
            {\author{H.}{Haken}}
    {Springer}{Berlin}{pp249--268}

\refartbook{refD109}
{\author{B.}{Candelpergher}, \author{F.}{Diener} \and
\author{M.}{Diener}}{1990}
{Retard \`a la bifurcation: du local au global}
            {Bifurcations of planar vector fields}
            {\author{J.P.}{Francoise} \and \author{R.}{Roussarie}}
            {Springer}{Berlin}{pp1--19}

\refarticle{refD110}
{\author{E.}{Beno\^it}, \author{F.}{Diener} \and
\author{M.}{Diener}}{1981}
{Chasse au canard}
            {Collectanea Mathematica}
            {31}{37--119}

\refbook{refD111}
{\author{P.}{Drazin}}{1992}
    {Nonlinear Systems}
    {Cambridge University Press}{Cambridge}

\refarticle{refE100}
{\author{A.}{Einstein}}{1905}
           {\"uber die von der molekular-kinetischen Theorie der
        W\"arme geforderte Bewegung von in ruhended Fl\"ussigkeiten}
           {Ann.Phys.}{17}{549-555}

\refartbook{refE101}
{\author{T.}{Erneux}, \author{E.C.}{Reiss}, \author{L.J.}{Holden}
\and \author{M.}{Georgiou}}{1991}
            {Slow Passage through bifurcation and limit points.
Asymptotic theory}
            {Dynamic Bifurcations}
            {\author{E.}{Beno\^\i t}}
            {Springer}{Berlin}{pp14--28}
 
\refarticletobe{refE102}
{\author{Stephen}{Einchcomb} \and \author{A.J.}{McKane}}{1993}
           {Quasi-monochromatic noise stuff with path integrals}
           {Phys. Rev. E}

\refartbook{refE103}
{\author{W.}{Eckhaus}}{1983}
{Relaxation oscillations including a standard chase on french ducks}
            {Asymptotic Analysis}
            {\author{F.}{Verhulst}}
            {Springer}{Berlin}{pp432--449}

\refarticle{refE104}
{\author{T.}{Erneux} \and \author{Paul}{Mandel}}{1984}
            {Slow passage through laser first threshold}
            {Phys. Rev. A}{39}{5179--5188}

\refarticle{refF5}
{\author{A.C.}{Fowler}, \author{J.D.}{Gibbon} \and
 \author{M.J.}{McGuinness}}{1983}
            {The real and complex Lorenz equations and their relevance to
             physical systems}
            {Physica}{7D}{126--134}
 
\refarticle{refF4}
{\author{A.C.}{Fowler}}{1990}
            {Homoclinic bifurcations in $n$ dimensions}
            {Studies in Applied Mathematics}{83}{193--209}

\refarticle{refF100}
{\author{L.}{Fronzoni}, \author{Frank}{Moss} \and
\author{P.V.E.}{McClintock}}{1987}
{Swept-parameter-induced postponements and noise on the Hopf bifurcation}
            {Phys. Rev. A}{36}{1492--1494}

\refarticle{refF101}
{\author{K.}{Fujimura} \and \author{R.E.}{Kelly}}{1994}
{Interaction between longitudinal convection rolls and transverse
waves in unstably stratified plane Poiseuille flow}
            {Phys. Fluids}{7}{25--65}

\refarticle{refF102}
{\author{Ronald Forrest}{Fox}}{1972}
{Contributions to the Theory of Multiplicative Stochastic Processes}
            {J. Math. Phys}{13}{1196--1207}
 
\refarticle{refF103}
{\author{Tadahisa}{Funaki}}{1983}
{Random motion of strings and related stochastic evolution equations}
            {Nagoya Math. J.}{89}{129--193}

\refarticle{refF104}
{\author{Tadahisa}{Funaki}}{1995}
{The scaling limit for a stochastic PDE and the separation of phases}
            {Probab. Theory Relat. Fields}{102}{221--288}

\refarticle{refG32}
{\author{P.}{Gaspard}, \author{R.}{Kapral} \and \author{G.}{Nicolis}}{1984}
            {Bifurcation phenomena near homoclinic systems: a two-parameter
             analysis}
            {J.~Stat. Phys.}{35}{697--727}

\refarticle{refG12}
{\author{P.}{Glendinning} \and \author{C.}{Sparrow}}{1984}
            {Local and global behaviour near homoclinic orbits}
            {J.~Stat. Phys.}{35}{645--696}
 
\refphd{refG22}
{\author{P.A.}{Glendinning}}{1985}
            {Homoclinic bifurcations}
            {University of Cambridge}
 
\refarticle{refG2}
{\author{P.}{Glendinning} \and \author{C.}{Sparrow}}{1986}
            {T-points: a codimension two heteroclinic bifurcation}
            {J.~Stat. Phys.}{43}{479--488}
 
\refarticle{refG45}
{\author{I.}{Goldhirsch}, \author{R.B.}{Pelz}, \author{S.A.}{Orszag}}{1989}
            {Numerical simulation of thermal convection in a two-dimensional
             finite box}
            {J.~Fluid Mech.}{199}{1--28}
 
\refbook{refG31}
{\author{D.}{Gottlieb} \and \author{S.A.}{Orszag}}{1977}
    {Numerical Analysis of Spectral Methods: Theory and Applications}
    {Society for Industrial and Applied Mathematics}{Philadelphia}
 
\refarticle{refG41}
{\author{E.}{Graham}}{1975}
            {Numerical simulation of two-dimensional compressible convection}
            {J.~Fluid Mech.}{70}{689--703}
 
\refarticle{refG42}
{\author{R.}{Graham}}{1976}
            {Onset of self-pulsing in lasers and the Lorenz model}
            {Phys. Lett.}{58A}{440--442}
 
\refarticle{refG43}
{\author{S.}{De~Gregorio}, \author{E.}{Scoppola} \and
 \author{B.}{Tirozzi}}{1983}
            {A rigorous study of periodic orbits by means of a computer}
            {J.~Stat. Phys.}{32}{25--33}
 
\refartbook{refG28}
{\author{J.}{Guckenheimer}}{1976}
            {A strange, strange attractor}
            {The Hopf Bifurcation and Its Applications}
            {\author{J.E.}{Marsden} \and \author{M.}{McCracken}}
            {Springer}{New York}{368--381}
 
\refarticle{refG30}
{\author{J.}{Guckenheimer} \and \author{E.}{Knobloch}}{1983}
            {Nonlinear convection in a rotating layer: amplitude expansions and
             normal forms}
            {Geophys. Astrophys. Fluid Dynamics}{23}{247--272}
 
\refbook{refG1}
{\author{J.}{Guckenheimer} \and \author{P.}{Holmes}}{1987}
            {Nonlinear Oscillations,
             Dynamical Systems and Bifurcations of Vector Fields}
            {Springer}{New York}

\refbook{refG100}
{\author{C.W.}{Gardiner}}{1990}
    {Handbook of Stochastic Methods}
    {Springer}{Berlin}

\refbook{refG101}
{\author{I.S.}{Gradshteyn} \and \author{M.}{Ryzhik}}{1980}
            {Table of integrals, series, and products} 
            {Academic Press}{New York}

\refartbook{refG102}
{\author{Paolo}{Grigolini}}{1989}
            {The projection approach to the Fokker-Planck equation:
application to phenomenological stochastic equations with colored noise}
            {Noise in nonlinear dynamical systems, Vol. I}
            {\author{Frank}{Moss} \and \author{P.V.E.}{McClintock}}
            {Cambridge University Press}{Cambridge}{pp161--190}

\refartbook{refG103}
{\author{Robert}{Graham}}{1989}
            {Macroscopic potentials, bifurcations and noise in
dissipative systems}
            {Noise in nonlinear dynamical systems, Vol. I}
            {\author{Frank}{Moss} \and \author{P.V.E.}{McClintock}}
            {Cambridge University Press}{Cambridge}{225--278}
  
\refarticle{refG104}
{\author{I.}{Gy\"ongy} \and \author{E.}{Pardoux}}{1993}
            {On quasi-linear stochastic partial differential equations}
            {Probab. Theory Relat. Fields}{94}{413--426}

\refarticle{refG105}
{\author{I.}{Gy\"ongy} \and \author{E.}{Pardoux}}{1993}
            {On the regularization effect of space-time white noise on
quasi-linear parabolic partial differential equations}
            {Probab. Theory Relat. Fields}{97}{211--230}

\refarticle{refG106}
{\author{R.}{Graham} \and \author{T.}{T\'el}}{1990}
            {Steady-state ensemble for the complex Ginzburg-Landau
equation with weak noise}
            {Phys. Rev. A}{42}{4661--4677}

\refarticle{refG107}
{\author{Hermann}{Grabert} \author{Peter}{H\ddot anggi}
\and \author{Peter}{Talkner}}{1979}
           {Is quantum mechanics equivalent to a classical stochastic process?}
            {Phys. Rev. A}{19}{2440--2445}

\refarticle{refG108}
{\author{J.}{Garc\'\i a-Ojalvo}
\and \author{J.M.}{Sancho}}{1994}
           {Colored noise in spatially extended systems}
            {Phys. Rev. E}{49}{2769--2778}

\refarticle{refG109}
{\author{J.}{Garc\'\i a-Ojalvo}, \author{J.M.}{Sancho}
\and \author{L.}{Ram\'\i rez-Piscina}}{1992}
           {Generation of spatiotemporal colored noise}
            {Phys. Rev. A}{46}{4670--4675}

\refarticle{refG110}
{\author{Robert}{Graham}}{1974}
            {Hydrodynamic fluctuations near the convection instability}
            {Phys. Rev. A}{10}{1762--1784}
\refarticle{refG111}
{\author{R.}{Graham} \and \author{A.}{Schenzle}}{1982}
            {Stabilization by multiplicative noise}
            {Phys. Rev. A}{26}{1676--1685}

\refarticle{refG112}
{\author{J.}{Garc\'\i a-Ojalvo}, \author{A.}{Hern\'andez-Machado} \and
\author{J.M.}{Sancho}}{1993} 
           {Effects of External Noise on the Swift-Hohenberg equation}
            {Phys. Rev. Lett.}{71}{1542--1545}

\refarticle{refG113}
{\author{P.}{Gaspard} et al (Ed.)}{1991}
            {Homoclinic Chaos}
            {Physica D}{62}{1--372}

\refarticle{refG114}
{\author{Raymind}{Goldstein} \and
\author{Gemunu H.}{Gunaratne}}{1991} 
           {Hydrodynamic and interfacial patterns with broken
space-time symmetry}
            {Phys. Rev. A}{43}{6700--6710}

\refarticle{refH13}
{\author{H.}{Haken}}{1975}
            {Analogy between higher instabilities in fluids and lasers}
            {Phys . Lett.}{53A}{77--78}
 
\refarticle{refH17}
{\author{S.M.}{Hammel}, \author{J.A.}{Yorke} \and \author{C.}{Grebogi}}{1988}
            {Numerical orbits of chaotic processes represent true orbits}
            {Bull. Am. Math. Soc.}{19}{465--469}
 
\refbook{refH15}
{\author{A.C.}{Hearne}}{1987}
            {Reduce User's Manual -- Version 3.3}
            {Rand}{Santa Monica}
 
\refarticle{refH3}
{\author{D.W.}{Hughes} \and \author{M.R.E.}{Proctor}}{1988}
            {Magnetic fields in the solar convection zone: magnetoconvection
             and magnetic buoyancy}
            {Ann. Rev. Fluid Mech.}{20}{187--223}
 
\refarticle{refH5}
{\author{H.E.}{Huppert} \and \author{D.R.}{Moore}}{1976}
            {Nonlinear double-diffusive convection}
            {J.~Fluid Mech.}{78}{821--854}
 
\refarticle{refH16}
{\author{N.E.}{Hurlburt} \and \author{J.}{Toomre}}{1988}
            {Magnetic fields interacting with nonlinear compressible convection}
            {Ap. J.}{327}{920--932}
 
\refarticle{refH1}
{\author{N.E.}{Hurlburt}, \author{M.R.E.}{Proctor},
 \author{N.O.}{Weiss} \and \author{D.P.}{Brownjohn}}{1989}
            {Nonlinear compressible magnetoconvection.
             Part~1. Travelling waves and oscillations}
            {J.~Fluid Mech.}{207}{587--628}

\refbook{refH100}
{\author{H.}{Haken(ed)}}{1981}
    {Chaos and Order in Nature}
    {Springer}{Berlin}
 
\refarticle{refH101}
{\author{D.W.}{Hughes} \and \author{M.R.E.}{Proctor}}{1990}
            {Chaos and the effect of noise in a model of three-wave
mode coupling}
            {Physica D}{46}{163--176}

\refarticle{refH102}
{\author{D.W.}{Hughes} \and \author{M.R.E.}{Proctor}}{1990}
            {A low order model of the shear instability of convection:
chaos and the effect of noise}
            {Nonlinearity}{3}{127--153}

\refarticle{refH103}
{\author{M.R.E.}{Proctor} \and\author{D.W.}{Hughes}}{1990}%
{The false Hopf bifurcation and noise sensitivity
        in bifurcations with symmetry}
            {Eur.J.Mech.,B/Fluids}{10}{81--86}

\refarticle{refH104}
{\author{Rebecca L.}{Honeycutt}}{1992}
            {Stochastic Runge-Kutta algorithms. I. White noise}
            {Phys. Rev. A}{45}{600--603}

\refbook{refH105}
{\author{Harry}{Hochstadt}}{1971}
    {The functions of mathematical physics}
    {Dover}{New York}

\refartbook{refH106}
{\author{Peter}{Hanggi}}{1989}
            {Colored noise in continuous dynamical systems: a
functional calculus approach}
            {Noise in nonlinear dynamical systems, Vol. I}
            {\author{Frank}{Moss} \and \author{P.V.E.}{McClintock}}
            {Cambridge University Press}{Cambridge}{307--328}
 
\refbook{refH107}
{\author{H.}{Haken}}{1978}
    {Synergetics}
    {Springer}{Berlin}

\refarticle{refH108}
{\author{Richard}{Haberman}}{1979}
            {Slowly varying jump and transition phenomena associated
with algebraic bifurcation problems}
            {SIAM J. Appl. Math.}{37}{69--106}

\refarticle{refH109}
{\author{J.J.}{Healy}, \author{D.S.}{Broomhead}, \author{K.A.}{Cliff},
\author{R.}{Jones} \and \author{T.}{Mullin}}{1991}
            {The origins of chaos in a modified Van der Pol oscillator}
            {Physica D}{48}{322--339}

\refbook{refH110}
{\author{W.}{Horsthemke} \& \author{R.}{Lefever}}{1984}
    {Noise-induced transitions}
    {Springer}{Berlin}
 
\refarticle{refH111}
{\author{W.}{Horsthemke} \and \author{R.}{Lefever}}{1977}
    {Phase transitions induced by external noise}
        {Phys. Lett. A}{64}{19--21}
 
\refarticle{refH112}
{\author{Michael R. E.}{Proctor} \and \author{Judith Y.}{Holyer}}{1986}
    {Planform selection in salt fingers}
        {J. Fluid Mech.}{168}{241--253}
\refarticle{refH113}
{\author{Lisa}{Holden} \and \author{Thomas}{Erneux}}{1993}
    {Slow passage through a Hopf bifurcation:
from oscillatory to steady state solutions}
        {SIAM J. Appl. Math.}{53}{1045--1058}
 
\refarticle{refH114}
{\author{P.}{Holmes}}{1979}
    {A nonlinear oscillator with a strange attractor}
        {Phil. Trans.}{292}{419--448}
\refarticle{refH115}
{\author{P.}{Holmes}}{1990}
    {Poincar\'e, celestial mechanics, dynamical-systems theory and `chaos'}
        {Phys. Rep.}{193}{137--163}
 
\refarticle{refH116}
{\author{L.N.}{Howard} \and \author{R.}{Krishnamurti}}{1986}
    {Large-scale flow in turbulent convection:
a mathematical model}
        {J. Fluid Mech.}{170}{385--410}

\refarticle{refH117}
{\author{P.C.}{Hohenberg} \and \author{J.W.}{Swift}}{1992}
            {Effects of additive noise at the onset of 
Rayleigh-B\'enard convection}
            {Phys. Rev. A}{46}{4773--4785}

\refarticle{refI100}
{\author{Kiyosi}{Ito}}{1964}
            {Expected number of zeros of continuous stationary Gaussian processes}
            {J. Math. Kyoto Univ.}{3-2}{207--216}
 
\refarticle{refJ100}
{\author{Peter}{Jung}, \author{George}{Gray}, \author{Rajarshi}{Roy}
 \and \author{Paul}{Mandel}}{1990}
    {Scaling Law for Dynamical Hysteresis}
        {Phys. Rev. Lett.}{65}{1873--1876}

\refphd{refJ5}
{\author{K.A.}{Julien}}{1991}
            {Strong spatial resonance in convection}
            {University of Cambridge}
 
\refarticle{refK2}
{\author{E.}{Knobloch} \and \author{M.R.E.}{Proctor}}{1981}
            {Nonlinear periodic convection in double-diffusive systems}
            {J.~Fluid Mech.}{108}{291--316}
 
\refarticle{refK5}
{\author{E.}{Knobloch}, \author{N.O.}{Weiss} \and \author{L.N.}{Da~Costa}}{1981}
            {Oscillatory and steady convection in a magnetic field}
            {J.~Fluid Mech.}{113}{153--186}
 
\refarticle{refK7}
{\author{E.}{Knobloch} \and \author{N.O.}{Weiss}}{1983}
            {Bifurcations in a model of magnetoconvection}
            {Physica}{9D}{379--407}
 
\refarticle{refK11}
{\author{E.}{Knobloch} \and \author{N.O.}{Weiss}}{1984}
            {Convection in sunspots and the origin of umbral dots}
            {Mon.~Not. R.~Astr. Soc.}{207}{203--214}
 
\refarticle{refK8}
{\author{E.}{Knobloch}}{1986}
            {On convection in a horizontal magnetic field with periodic
             boundary conditions}
            {Geophys. Astrophys. Fluid Dynamics}{36}{161--177}
 
\refarticle{refK19}
{\author{E.}{Knobloch}}{1986}
            {On the degenerate Hopf bifurcation with $O(2)$ symmetry}
            {Contemp. Math.}{56}{193--201}
 
\refarticle{refK24}
{\author{E.}{Knobloch}}{1986}
            {Normal forms for bifurcations at a double zero eigenvalue}
            {Phys. Lett.}{115A}{199--201}
 
\refarticle{refK8a}
{\author{E.}{Knobloch}}{1986a}
            {On convection in a horizontal magnetic field with periodic
             boundary conditions}
            {Geophys. Astrophys. Fluid Dynamics}{36}{161--177}
 
\refarticle{refK19b}
{\author{E.}{Knobloch}}{1986b}
            {On the degenerate Hopf bifurcation with $O(2)$ symmetry}
            {Contemp. Math.}{56}{193--201}
 
\refarticle{refK24c}
{\author{E.}{Knobloch}}{1986c}
            {Normal forms for bifurcations at a double zero eigenvalue}
            {Phys. Lett.}{115A}{199--201}
 
\refarticle{refK4}
{\author{E.}{Knobloch}, \author{D.R.}{Moore},
 \author{J.}{Toomre} \and \author{N.O.}{Weiss}}{1986}
            {Transitions to chaos in two-dimensional double-diffusive
             convection}
            {J.~Fluid Mech.}{166}{409--448}
 
\refarticle{refK1}
{\author{E.}{Knobloch} \and \author{M.R.E.}{Proctor}}{1988}
            {The double Hopf bifurcation with $2:1$ resonance}
            {Proc. R.~Soc. Lond.}{415A}{61--90}
 
\refarticle{refK17}
{\author{E.}{Knobloch} \and \author{M.}{Silber}}{1990}
            {Travelling wave convection in a rotating layer}
            {Geophys. Astrophys. Fluid Dynamics}{51}{195--209}

\refbook{refK34}
{\author{D.E.}{Knuth}}{1984}
            {The \TeX book}
            {Addison Wesley}{Reading}
\refbook{refK100}
{\author{Peter E.}{Kloeden} \and \author{Eckhard}{Platen}}{1992}
    {Numerical Solution of Stochastic Differential Equations}
    {Springer}{Berlin}

\refbook{refK101}
{\author{H.}{Kunita}}{1990}
    {Stochastic Flows and Stochastic Differential Equations}
    {Cambridge University Press}{Cambridge}

\refbook{refK102}
{\author{T.W.}{K\ddot orner}}{1988}
    {Fourier analysis}
    {Cambridge University Press}{Cambridge}

\refarticle{refK103}
{\author{N.}{van Kampen}}{1981}
            {\ito\ Versus Stratonovich}
            {J. Stat. Phys.}{24}{175--187}
\refarticle{refK104}
{\author{D.K.}{Kondepudi} \and \author{G.W.}{Nelson}}{1985}
            {Weak neutral currents and the origin of biomolecular chirality}
            {Nature}{314}{438--441}
\refarticle{refK105}
{\author{D.K.}{Kondepudi}, \author{F.}{Moss} \and 
\author{P.V.E.}{McClintock}}{1986}
            {Observation of symmetry breaking, state selection and
sensitivity in a noisy electronic system}
            {Physica D}{21}{296--306}
 
\refbook{refK106}
{\author{Frank B.}{Knight}}{1981}
    {Essentials of Brownian motion and diffusion}
    {American Mathematical Society}{Providence}

\refarticle{refK107}
{\author{Raymond}{Kapral} \and \author{Paul}{Mandel}}{1985}
            {Bifurcation structure of the nonautonomous quadratic map}
            {Phys. Rev. A}{32}{1076--1081}

\refbook{refK108}
{\author{J.}{Kevorkian} \and \author{J.D.}{Cole}}{1981}
    {Perturbation Methods In Applied Mathematics}
    {Springer}{New York}

\refarticle{refK109}
{\author{D.J.}{Kaup}, \author{A.}{Rieman} \and \author{A.}{Bers}}{1979}
            {Space-time evolution of nonlinear three-wave interactions}
            {Rev. Mod. Phys.}{51}{275--309}
 
\refbook{refK110}
{\author{Ioannis}{Karatzas} \and \author{Steven E.}{Shreve}}{1988}
    {Brownian Motion and Stochastic Calculus}
    {Springer}{New York}

\refarticle{refL5}
{\author{E.N.}{Lorenz}}{1963}
            {Deterministic nonperiodic flow}
            {J.~Atmos. Sci.}{20}{130--141}
 
\refarticle{refL12}
{\author{D.V.}{Lyubimov} \and \author{M.A.}{Zaks}}{1983}
            {Two mechanisms of the transition to chaos in finite-dimensional
             models of convection}
            {Physica}{9D}{52--64}
 
\refarticle{refL18}
{\author{D.V.}{Lyubimov}, \author{A.S.}{Pikovsky} \and
 \author{M.A.}{Zaks}}{1989}
            {Universal scenarios of transition to chaos via homoclinic
             bifurcations}
            {Sov. Sci. Rev.~C. Math. Phys.}{8}{221--292}

\refarticle{refL23}
{\author{A.S.}{Landsberg} \and \author{E.}{Knobloch}}{1991}
            {Direction-reversing traveling waves}
            {Phys. Lett. A}{159}{17--20}

\refarticle{refL24}
{\author{A.S.}{Landsberg} \and \author{E.}{Knobloch}}{1993}
            {New types of waves in systems with $O(2)$ symmetry}
            {Phys. Lett. A}{179}{316--324}

\refarticle{refL100}
{\author{G.D.}{Lythe} \and \author{M.R.E.}{Proctor}}{1993}
            {Noise and slow-fast dynamics in a three-wave resonance problem}
            {Phys. Rev. E. }{47}{3122-3127}
 
\refarticle{refL101}
{\author{M.R.E.}{Proctor} \and \author{G.D.}{Lythe}}{1993}
            {Noise and resonant mode interactions}
            {Ann. New York Acad. Sci. }{706}{42-53}

\refartbook{refL102}
{\author{Claude}{Lobry}}{1991}
            {Dynamic Bifurcations}
            {Dynamic Bifurcations}
            {\author{E.}{Beno\^\i t}}
            {Springer}{Berlin}{pp1--13}
 
\refartbook{refL103}
{\author{Katya}{Lindenberg}, {Bruce J.}{West} \and {Jaume}{Masoliver}}{1989}
            {First passage time problems for non-Markovian processes}
            {Noise in nonlinear dynamical systems, Vol. I}
            {\author{Frank}{Moss} \and \author{P.V.E.}{McClintock}}
            {Cambridge University Press}{Cambridge}{110--160}
 
\refbook{refL104}
{\author{L.}{Landau} \and \author{E.}{Lifshitz}}{1959}
    {Statistical Physics}
    {Pergamon}{London}

\refarticle{refL105}
{\author{N.R.}{Lebovitz} \and \author{R.J.}{Schaar}}{1975}
            {Exchange of stabilities in autonomous systems}
            {Studies in Appl. Math. }{54}{229--260}

\refarticle{refL106}
{\author{Claude}{Lobry}}{1992}
        {A propos du sens des textes math\'ematiques. Un example:
        la th\'eorie des ``bifurcations dynamiques''}
            {Ann. Inst. Fourier}{42}{327-352}

 
\refartbook{refL109}
{\author{G.D.}{Lythe}}{1994}
            {A noise-controlled dynamic bifurcation}
            {Chaos and Nonlinear Mechanics}
                {\author{T.}{Kapitaniak} and \author{J.}{Brindley}(eds)}
{World Scientific}{Singapore}{147--158}
 
\refbook{refL110}
{\author{Andrej}{Lasota} \and \author{Michael C.}{Mackey}}{1985}
    {Probabilistic properties of deterministic systems}
    {Cambridge University Press}{Cambridge}

\refarticle{refL111}
{\author{Pui-Man}{Lam} \and \author{Diola}{Bagayoko}}{1993}
            {Spatiotemporal correlation of coloured noise}
            {Phys. Rev. E}{48}{3267--3270}
\refarticle{refL112}
{\author{Katya}{Lindenberg}, {Bruce J.}{West} \and {Raoul}{Kopelman}}{1989}
            {Steady-State Segregation in Diffusion-Limited Reactions}
            {Phys. Rev. Lett.}{60}{1777--1780}

\refbook{refL113}
{\author{Leon}{Lapidus} \and \author{George F.}{Pinder}}{1982}
    {Numerical Solution of Partial Differential Equations in Science and Engineering}
    {Wiley}{New York}

\refartbook{refL114}
{\author{G.D.}{Lythe}}{1994}
            {Noise and Dynamic transitions}
            {Stochastic Partial Differential Equations}
                {\author{Alison}{Etheridge}}{Cambridge University Press}
                {Cambridge}{}

\refarticle{refL115}
{\author{G.D.}{Lythe}}{1995}
           {Dynamics controlled by additive noise}
           {Nuovo Cimento D}{17}{855--861}

\refphd{refL116}
{\author{G.D.}{Lythe}}{1994}
            {Stochastic slow-fast dynamics}
            {University of Cambridge}

\refarticletobe{refL117}
{\author{C.-H.}{Lu} \and \author{R.M.}{Evan-Iwanowski}}{1994}
           {The Nonstationary Effects on the Logistic Map and the
Softening Duffing Oscillator}
           {}

\refartbook{refM28}
{\author{B.A.}{Malomed} \and \author{A.A.}{Nepomnyashchy}}{1990}
            {Onset of chaos in the generalized Ginzburg--Landau equation}
            {Nonlinear Evolution of Spatio-Temporal Structures in Dissipative
             Continuous Systems}
            {\author{F.H.}{Busse} \and \author{L.}{Kramer}}
            {Plenum Press}{New York}{419--424}

\refarticle{refM25}
{\author{W.V.R.}{Malkus} \and \author{G.}{Veronis}}{1958}
            {Finite amplitude cellular convection}
            {J.~Fluid Mech.}{4}{225--260}
 
\refarticle{refM19}
{\author{P.S.}{Marcus}}{1981}
            {Effects of truncation in modal representations of thermal
             convection}
            {J.~Fluid Mech.}{103}{241--255}
 
\refartbook{refM28}
{\author{B.A.}{Malomed} \and \author{A.A.}{Nepomnyashchy}}{1990}
            {Onset of chaos in the generalized Ginzburg--Landau equation}
            {Nonlinear Evolution of Spatio-Temporal Structures in Dissipative
             Continuous Systems}
            {\author{F.H.}{Busse} \and \author{L.}{Kramer}}
            {Plenum Press}{New York}{419--424}
 
\refbook{refM34}
{\author{P.}{Manneville}}{1990}
            {Dissipative Structures and Weak Turbulence}
            {Academic Press}{Boston}
 
\refarticle{refM5}
{\author{R.M.}{May}}{1976}
            {Simple mathematical models with very complicated dynamics}
            {Nature}{261}{459--467}
 
\refarticle{refM20}
{\author{D.R.}{Moore} \and \author{N.O.}{Weiss}}{1973}
            {Two-dimensional Rayleigh--B\'enard convection}
            {J.~Fluid Mech.}{58}{289--312}
 
\refarticle{refM22}
{\author{D.R.}{Moore}, \author{N.O.}{Weiss} \and \author{J.M.}{Wilkins}}{1990}
            {The reliability of numerical experiments:
             transitions to chaos in thermosolutal convection}
            {Nonlinearity}{3}{997--1014}

\refarticle{refM100}
{\author{Paul}{Mandel} \and \author{T.}{Erneux}}{1984}
            {Laser Lorenz equations with a time-dependent parameter}
            {Phys. Rev. Lett.}{53}{1818--1820}

\refarticle{refM101}
{\author{Paul}{Mandel} \and \author{Thomas}{Erneux}}{1987}
            {The slow passage through a steady bifurcation:
        delay and memory effects}
            {J. Stat. Phys.}{48}{1059--1070}

\refarticle{refM102}
{\author{R.}{Mannella}, \author{Frank}{Moss}
\and \author{P.V.E.}{McClintock}}{1987}
           {Postponed bifurcations of a ring-laser with a swept parameter
        and additive colored noise}
           {Phys. Rev. A}{35}{2560--2566}
\refbook{refM103}
{\author{Frank}{Moss} \and \author{P.V.E.}{McClintock}}{1987}
            {Noise and nonlinear dynamical systems}
            {Cambridge University Press}{Cambridge}

\refarticle{refM104}
{\author{Miltiades}{Georgiou} \and \author{Thomas}{Erneux}}{1992}
          {Pulsating laser oscillations depend on 
        extremely-small-amplitude noise}
           {Phys. Rev. A}{45}{6636--6642}

\refbook{refM105}
{\author{A}{Atchison}}{19**}
            {The lognormal distribution}
            {Cambridge University Press}{Cambridge}

\refarticletobe{refM106}
{\author{Michael C.}{Mackey} \and \author{Irina G.}{Nechaeva}}{1993}
            {Noise and stability in differential delay equations}
             {Journal of Dynamics and Differential Equations}

\refarticle{refM107}
{\author{Christopher W.}{Meyer}, \author{Guenter}{Ahlers}
\and \author{David S.}{Cannell}}{1991}
            {Stochastic influences on pattern formation in
Rayleigh-B\'enard convection: Ramping experiments}
             {Phys. Rev. A}{44}{2514--2537}

\refbook{refM108}
{\author{Kenneth S.}{Miller}}{1974}
            {Complex Stochastic Processes}
            {Addison-Wesley}{Reading}
\refbook{refM208}
{\author{Kenneth S.}{Miller}}{1964}
            {Multidimensional Gaussian Distributions}
            {Wiley}{New York}

\refarticle{refM109}
{\author{R.}{Mannella}
\and \author{P.V.E.}{McClintock}}{1990}
           {Noise in nonlinear dynamical systems}
           {Contemporary Physics}{31}{179--194}

\refarticle{refM110}
{\author{P.C.}{Matthews}, \author{M.R.E.}{Proctor},
\author{A.M.}{Rucklidge} \and \author{N.O.}{Weiss}}{1993}
            {Pulsating waves in nonlinear magnetoconvection}
            {Phys. Lett. A}{183}{69--75}
 
\refarticle{refM111}
{\author{Frank}{Moss}, \author{D.K.}{Kondepudi}
 \and \author{P.V.E.}{McClintock}}{1985}
            {Branch selectivity at the bifurcation of a bistable
system with external noise}
            {Phys. Lett. A}{112}{293--296}
 
\refarticle{refM112}
{\author{Brian}{Morris} \and \author{Frank}{Moss}}{1986}
            {Postponed bifurcations of a quadratic map with a swept parameter}
            {Phys. Lett. A}{118}{117--120}
 
\refarticle{refM113}
{\author{C.}{Meunier} \and \author{A.D.}{Verga}}{1988}
            {Noise and Bifurcations}
            {J. Stat. Phys.}{50}{345--375}
 
\refarticle{refM114}
{\author{Michael C.}{Mackey}, \author{Andr\'e}{Longtin}
\and \author{Andrzej}{Lasota}}{1990}
            {Noise-Induced Global Asymptotic Stability}
             {J. Stat. Phys.}{60}{735--751}

\refarticle{refN2}
{\author{R.S.}{Mckay} \and \author{C.}{Tresser}}{1987}
            {Some flesh on the skeleton: the bifurcation structure
                of bimodal maps}
            {Physica}{27D}{412--422}

\refarticle{refN3}
{\author{M.}{Nagata}, \author{M.R.E.}{Proctor} \and
 \author{N.O.}{Weiss}}{1990}
            {Transitions to asymmetry in magnetoconvection}
            {Geophys. Astrophys. Fluid Dynamics}{51}{211--241}
 
\refarticle{refN2}
{\author{A.C.}{Newell} \and \author{J.A.}{Whitehead}}{1969}
            {Finite bandwidth, finite amplitude convection}
            {J.~Fluid Mech.}{38}{279--303}
 
\refarticle{refN4}
{\author{J.}{Niederl\"ander}, \author{M.}{L\"ucke} \and
 \author{M.}{Kamps}}{1991}
            {Weakly nonlinear convection: Galerkin model, numerical simulation,
             and amplitude equation}
            {Z.~Phys.~B -- Condensed Matter}{82}{135--141}

\refarticle{refN100}
{\author{A.I.}{Neishtadt}}{1987}
           {Persistence of stability loss for dynamical bifurcations}
           {Differential Equations}{23}{1385--1391}

\refarticle{refN101}
{\author{Bruce}{McNamara} and {Kurt}{Wiesenfeld}}{1989}
           {Theory of stochastic resonance}
           {Phys. Rev. A}{39}{4854--4869}

\refarticle{refN102}
{\author{Alan C.}{Newell}, \author{Thierry}{Passot}
 \and \author{Joceline}{Lega}}{1993}
            {Order parameter equations for patterns}
            {Ann. Rev. Fluid Mech.}{25}{399--453}

\refarticle{refN103}
{\author{Ali H.}{Nayfeh} \and \author{Nestor E.}{Sanchez}}{1989}
            {Bifurcations in a forced softening Duffing oscillator}
            {Int. J. Non-Linear Mechanics}{24}{483--497}
 
\refarticle{refN104}
{\author{S.}{Novak} \and \author{R. G.}{Frehlich}}{1982}
            {Transition to chaos in the Duffing oscillator}
            {Phys. Rev. A}{26}{3660--3663}
 
\refbook{refO100}
{\author{B.}{\O ksendal}}{1989}
    {Stochastic Differential Equations}
    {Springer}{Berlin}

\refbook{refP6}
{\author{T.S.}{Parker} \and \author{L.O.}{Chua}}{1989}
            {Practical Numerical Algorithms for Chaotic Systems}
            {Springer}{New York}
 
\refbook{refP2}
{\author{W.H.}{Press}, \author{B.P.}{Flannery},
 \author{S.A.}{Teukolsky} \and \author{W.T.}{Vetterling}}{1986}
            {Numerical Recipes -- the Art of Scientific Computing}
            {Cambridge University Press}{Cambridge}
 
\refarticle{refP4}
{\author{M.R.E.}{Proctor} \and \author{C.A.}{Jones}}{1988}
            {The interaction of two spatially resonant patterns in thermal
             convection. Part~1. Exact $1:2$~resonance}
            {J.~Fluid Mech.}{188}{301--335}
 
\refarticle{refP1}
{\author{M.R.E.}{Proctor} \and \author{N.O.}{Weiss}}{1982}
            {Magnetoconvection}
            {Rep. Prog. Phys.}{45}{1317--1379}
 
\refarticle{refP8}
{\author{M.R.E.}{Proctor} \and \author{N.O.}{Weiss}}{1990}
            {Normal forms and chaos in thermosolutal convection}
            {Nonlinearity}{3}{619--637}

\refartbook{refP100}
{\author{M.R.E.}{Proctor} \and \author{D.W.}{Hughes}}{1990}
        {Chaos and the effect of noise for the double Hopf
bifurcation with 2:1 resonance}
        {Nonlinear Evolution of Spatio-Temporal Structures in
Dissipative Continuous Systems}
        {F H Busse and L Kramer}
        {Plenum}{New York}{}
\refarticle{refP101}
{\author{F.}{de Pasquale}, \author{J.M.}{Sancho}, \author{M.}{San Miguel}
 \and \author{P.}{Tartaglia}}{1986}
            {Decay of an unstable state in the presence of
multiplicative noise}
            {Phys. Rev. A}{33}{4360--4366}

\refarticle{refP102}
{\author{P.}{Piera\'nski}
 \and \author{J.}{Ma\l ecki}}{1987}
            {Noise-sensitive Hysteresis Loops around Period-Doubling
Bifurcation Points}
            {Il Nuovo Cimento}{9}{757--779}
 
\refbook{refP103}
{\author{Guiseppe}{Da Prato} \and \author{Jerzy}{Zabczyk}}{1992}
    {Stochastic Equations in Infinite Dimensions}
    {Cambridge University Press}{Cambridge}     

\refbook{refP104}
{\author{Phoolan}{Prasad} \and \author{Renuka}{Ravindran}}{1985}
    {Partial Differential Equations}
    {Wiley Eastern}{New Delhi}  

\refarticle{refP105}
{\author{Nathan}{Platt}, \author{Stephen M.}{Hammel}
\and \author{James F.}{Heagy}}{1994}
            {Effect of Additive Noise on On-Off Intermittency}
            {Phys. Rev. Lett.}{52}{3498--3501}

\refbook{refP106}
{\author{Philip}{Protter}}{1990}
    {Stochastic Integration and Differential Equations}
    {Springer}{Berlin}

\refarticle{refR5}
{Lord~Rayleigh}{1916}
            {On convection currents in a horizontal layer of fluid, when the
             higher temperature is on the under side}
            {Phil.~Mag.}{32}{529--546}
 
\refarticle{refR13}
{\author{A.J.}{Rodr\'\i guez-Luis}, \author{E.}{Freire}
 \and \author{E.}{Ponce}}{1991}
            {On a codimension 3 bifurcation arising in an autonomous electronic
             circuit}
            {International Series of Numerical Mathematics}{97}{301--306}
 
\refarticle{refR8}
{\author{H.T.}{Rossby}}{1969}
            {A study of B\'enard convection with and without rotation}
            {J.~Fluid Mech.}{36}{309--335}
 
\refarticle{refR2}
{\author{A.}{Rucklidge} \and \author{S.}{Zaleski}}{1988}
            {A microcanonical model for interface formation}
            {J.~Stat. Phys.}{51}{299--307}
 
\refarticle{refR9}
{\author{A.M.}{Rucklidge}}{1992}
            {Chaos in models of double convection}
            {J.~Fluid Mech.}{237}{209--229}



\refbook{refR101}
{\author{Daniel}{Revuz} \and \author{Marc}{Yor}}{1991}
            {Continuous Martingales and Brownian Motion}
            {Springer}{Berlin}

\refbook{refR100}
{\author{H.}{Risken}}{1984}
            {The Fokker-Planck Equation}
            {Springer}{Berlin}
\refarticle{refR102}
{\author{L.A.}{Rubenfeld}}{1979}
            {A model bifurcation problem exhibiting the effects of
slow passage through critical}
            {SIAM J. Appl. Math.}{37}{302--306}
 
\refphd{refR103}
{\author{A.M.}{Rucklidge}}{1991}
            {Chaos in models of double convection}
            {University of Cambridge}
  
\refarticle{refR105}
{\author{A.M.}{Rucklidge} \and \author{P.C.}{Matthews}}{1995}
            {Analysis of the shearing instability
 in nonlinear magnetoconvection}
            {Nonlinearity}{9}{311--352}

\refarticle{refS31}
{\author{T.}{Sauer} \and \author{J.A.}{Yorke}}{1991}
            {Rigorous verification of trajectories for the computer simulation
             of dynamical systems}
            {Nonlinearity}{4}{961--979}
 
 

\refarticle{refS13}
{\author{L.P.}{Shil'nikov}}{1965}
            {A case of the existence of a countable number of periodic motions}
            {Soviet Maths. Dokl.}{6}{163--167}

\refarticle{refS25}
{\author{T.}{Shimizu} \and \author{N.}{Morioka}}{1978}
            {Chaos and limit cycles in the Lorenz model}
            {Phys. Lett.}{66A}{182--184}
 
\refarticle{refS18}
{\author{T.}{Shimizu} \and \author{N.}{Morioka}}{1980}
            {On the bifurcation of a symmetric limit cycle to an
             asymmetric one in a simple model}
            {Phys. Lett.}{76A}{201--204}

\refarticle{refS10}
{\author{T.G.L.}{Shirtcliffe}}{1969}
            {An experimental investigation of thermosolutal convection at
             marginal stability}
            {J.~Fluid Mech.}{35}{677--688}
 
\refarticle{refS19}
{\author{L.}{Sirovich} \and \author{A.E.}{Deane}}{1991}
            {A computational study of Rayleigh--B\'enard convection. Part~2.
             Dimension considerations}
            {J.~Fluid Mech.}{222}{251--265}
 
\refbook{refS1}
{\author{C.}{Sparrow}}{1982}
            {The Lorenz Equations: Bifurcations, Chaos, and Strange Attractors}
            {Springer}{New York}
 
\refarticle{refS9}
{\author{E.A.}{Spiegel}}{1987}
            {Chaos: a mixed metaphor for turbulence}
            {Proc. R.~Soc. Lond.}{413A}{87--95}
 
\refarticle{refS20}
{\author{J.W.}{Swift} \and \author{K.}{Wiesenfeld}}{1984}
            {Suppression of period doubling in symmetric systems}
            {Phys. Rev. Lett.}{52}{705--708}
 
\refphd{refS30}
{\author{J.}{Swinton}}{1991}
            {Stability of homoclinic waves in lasers and fluids}
            {Imperial College, London}
 
\refarticle{refS6}
{\author{P.}{Sz\'epfalusy} \and \author{T.}{T\'el}}{1985}
            {Properties of maps related to flows around a saddle point}
            {Physica}{16D}{252--264}
\refarticle{refS100}
{\author{Emily}{Stone} \and \author{Philip}{Holmes}}{1990}
            {Random perturbations of heteroclinic attractors}
            {SIAM J. Appl. Math.}{50}{726--743}

\refbook{refS101}
{\author{R}{Stratonovich}}{1963}
            {Topics in the theory of random noise}
            {Gordon and Breach}{New York}

\refarticle{refS102}
{\author{N.G.}{Stocks}, \author{R.}{Mannella} \and \author{P.V.E.}{McClintock}}{1989}
            {Influence of random fluctuations on delayed bifurcations:
        The case of additive white noise}
            {Phys. Rev. A}{40}{5361--5369}

\refarticle{refS103}
{\author{N.G.}{Stocks}, \author{R.}{Mannella} \and \author{P.V.E.}{McClintock}}{1990}
            {Influence of random fluctuations on delayed bifurcations.
        II. The cases of white and coloured additive 
        and multiplicative noise}
            {Phys. Rev. A}{42}{3356--3362}

\refartbook{refS104}
{\author{R.L.}{Stratonovich}}{1989}
            {Some Markov methods in the theory of stochastic processes
in nonlinear dynamical systems}
            {Noise in nonlinear dynamical systems, Vol. I}
            {\author{Frank}{Moss} \and \author{P.V.E.}{McClintock}}
            {Cambridge University Press}{Cambridge}{16--71}
 
\refartbook{refS105}
{\author{J.M.}{Sancho} \and \author{M.}{San Miguel}}{1989}
            {Langevin equations with coloured noise}
            {Noise in nonlinear dynamical systems, Vol. I}
            {\author{Frank}{Moss} \and \author{P.V.E.}{McClintock}}
            {Cambridge University Press}{Cambridge}{72--109}
 
\refarticle{refS106}
{\author{N.G.}{Stocks}, \author{N.D.}{Stein} \and \author{P.V.E.}{McClintock}}{1993}
            {Stochastic resonance in monostable systems}
            {J. Phys. A}{26}{L385--L390}

\refarticle{refS107}
{\author{David}{Sigeti} \and \author{Werner}{Horsthemke}}{1989}
            {Pseudo-regular oscillations induced by external noise}
            {J. Stat. Phys}{50}{1217--1222}

\refarticle{refS108}
{\author{Steven H.}{Strogatz}, \author{Renato E.}{Mirollo}
\and \author{Paul C.}{Matthews}}{1992}
            {Coupled Nonlinear Oscillators below the Synchronization
Threshold: Relaxation by Generalized Landau Damping}
            {Phys. Rev. Lett.}{68}{2730--2733}

\refarticle{refS109}
{\author{J.W.}{Swift}, \author{P.C.}{Hohenberg}
 \and \author{Guenter}{Ahlers}}{1991}
            {Stochastic Landau equation with time-dependent drift}
            {Phys. Rev. A}{43}{6572--6580}
 
\refarticle{refS110}
{\author{John}{Smythe}, \author{Frank}{Moss} \and
\author{P.V.E.}{McClintock}}{1983}
{Observation of a Noise-induced Phase Transition with an Analog Simulator}
            {Phys. Rev. Lett.}{51}{1062--1065}

\refarticle{refS111}
{\author{V.E.}{Shapiro}}{1993}
            {Systems near a critical point under multiplicative noise 
and the concept of effective potential}
            {Phys. Rev. E}{48}{109--120}

\refarticle{refS112}
{\author{V.E.}{Shishkova}}{1973}
            {Examination of a system of differential equations with a
small parameter in the highest derivatives}
            {Soviet Math. Dokl.}{14}{183--187}

\refbook{refS113}
{\author{Joel}{Smoller}}{1983}
            {Shock Waves and Reaction-Diffusion Equations}
            {Springer}{New York}

\refarticle{refS114}
{\author{D.J.}{Scalapino}, \author{M.}{Sears} \and
\author{R.A.}{Ferrell}}{1972}
        {Statistical Mechanics of One-Dimensional Ginzburg-Landau Fields}
         {Phys. Rev. B}{6}{3409--3416}

\refarticle{refS115}
{\author{Masuo}{Suzuki}}{1978}
        {Theory of instability, Brownian motion and formation of
macroscopic order}
         {Phys. Lett. A}{67}{339--341}
 
\refarticle{refT4}
{\author{F.}{Takens}}{1974}
            {Forced oscillations and bifurcations}
            {Comm. Math. Inst., Rijksuniversiteit Utrecht}{3}{1--59}
 
\refarticle{refV1}
{\author{G.}{Veronis}}{1965}
            {On finite amplitude instability in thermohaline convection}
            {J.~Marine Res.}{23}{1--17}
 
\refarticle{refV4}
{\author{G.}{Veronis}}{1966}
            {Motions at subcritical values of the Rayleigh number in a rotating
             fluid}
            {J.~Fluid Mech.}{24}{545--554}
 
\refarticle{refV7}
{\author{G.}{Veronis}}{1966}
            {Large-amplitude B\'enard convection}
            {J.~Fluid Mech.}{26}{49--68}
 
\refarticle{refV4a}
{\author{G.}{Veronis}}{1966a}
            {Motions at subcritical values of the Rayleigh number in a rotating
             fluid}
            {J.~Fluid Mech.}{24}{545--554}
 
\refarticle{refV7b}
{\author{G.}{Veronis}}{1966b}
            {Large-amplitude B\'enard convection}
            {J.~Fluid Mech.}{26}{49--68}
 
\refarticle{refV5}
{\author{G.}{Veronis}}{1968}
            {Large-amplitude B\'enard convection in a rotating fluid}
            {J.~Fluid Mech.}{31}{113--139}

\refarticle{refW12}
{\author{R.W.}{Walden}, \author{P.}{Kolodner},
 \author{A.}{Passner} \and \author{C.M.}{Surko}}{1985}
            {Travelling waves and chaos in convection in binary fluid mixtures}
            {Phys. Rev. Lett.}{55}{496--499}
 
\refarticle{refW16}
{\author{N.O.}{Weiss}}{1966}
            {The expulsion of magnetic flux by eddies}
            {Proc. Roy. Soc.}{293A}{310--328}
 
\refarticle{refW3}
{\author{N.O.}{Weiss}}{1981}
            {Convection in an imposed magnetic field. Part 1. The development
             of nonlinear convection}
            {J.~Fluid Mech.}{108}{247--272}
 
\refarticle{refW4}
{\author{N.O.}{Weiss}}{1981}
            {Convection in an imposed magnetic field. Part~2. The dynamical
             regime}
            {J.~Fluid Mech.}{108}{273--289}
 
\refarticle{refW3a}
{\author{N.O.}{Weiss}}{1981a}
            {Convection in an imposed magnetic field. Part~1. The development
             of nonlinear convection}
            {J.~Fluid Mech.}{108}{247--272}
 
\refarticle{refW4b}
{\author{N.O.}{Weiss}}{1981b}
            {Convection in an imposed magnetic field. Part~2. The dynamical
             regime}
            {J.~Fluid Mech.}{108}{273--289}
 
\refarticle{refW11}
{\author{N.O.}{Weiss}, \author{D.P.}{Brownjohn},
 \author{N.E.}{Hurlburt} \and \author{M.R.E.}{Proctor}}{1990}
            {Oscillatory convection in sunspot umbrae}
            {Mon.~Not. R.~Astr. Soc.}{245}{434--452}

\refbook{refW5}
{\author{S.}{Wiggins}}{1988}
    {Global Bifurcations and Chaos: Analytical Methods}
    {Springer}{New York}
 
\refarticle{refV100}
{\author{S.Ya.}{Vyshkind} \and \author{M.I.}{Rabinovich}}{1976}
            {The phase stochastization mechanism and the structure of
wave turbulence in dissipative media}
            {Sov. Phys. JETP}{44}{292--299}
\refarticle{refW100}
{\author{J.-M.}{Wersinger}, \author{J.M.}{Finn}
  \and \author{E.}{Ott}}{1980}
            {Bifurcation and ``strange'' behaviour in instability
saturation by nonlinear three-wave mode coupling}
            {Phys. Fluids}{23}{1142--1154}

\refnote{refNote1}
{The change of variables from the equations for
the amplitudes of the three wave modes used here differs  from
that of Hughes and Proctor; our $(x,y,z)$ corresponds to $(-x,w,y)$}

\refnote{refNote2}
{The distribution of a variable whose logarithm is normally distributed is 
sometimes called lognormal.
In an earlier publication[3], we were unaware of
this convention and used the term `lognormal` for the distribution of
the logarithm of $\ha \ln(n^2)$
 where $n$ is a Gaussian random variable with unit variance.}

\refnote{refNote3}{In an earlier publication~[12] we called this the
``log-normal'' distribution, being unaware that that term is
conventionally used for the distribution of a variable whose logarithm
is normally distributed}

\refnote{refNote4}{Hughes and Proctor (reference [5]) performed a
similar reduction. The variables $x,y,z$ used here are related to the
$A,B,C$ used by them via $x=A$, $y=\ha(B+C)$ and $z=\ha(B-C)$; $\delta
= \ha(1-\lambda)$. This choice of variables used in this work
simplifies analysis because of the symmetry (6)}

\refarticle{refT100}
{\author{M.C.}{Torrent} \and \author{M.}{San Miguel}}{1988}
           {Stochastic-dynamics characterization of delayed laser
threshold instability with swept control parameter}
           {Phys. Rev. A}{38}{245--251}

\refarticle{refT101}
{\author{M.C.}{Torrent}, \author{F.}{Sagu\'es} \and \author{M.}{San Miguel}}{1989}
           {Dynamics of sweeping through an instability: Passage-time
statistics for colored noise}
           {Phys. Rev. A}{40}{6662--6672}

\refarticle{refT102}
{\author{M.}{Tlidi}, \author{Paul}{Mandel}  \and
\author{R.}{Lefever}}{1994}
           {Localized Structures and Localized Patterns in Optical Bistability}
           {Phys. Rev. Lett}{73}{640--643}

\refbook{refW101}
{\author{David}{Williams}}{1991}
            {Probability with Martingales}
            {Cambridge University Press}{Cambridge}

\refarticle{refV100}
{\author{Jorge}{Vi\~nals}, \author{Hao-Wen}{Xi}
  \and \author{J.D.}{Gunton}}{1992}
            {Numerical study of the influence of forcing terms
and fluctuations near onset on the roll pattern in Rayleigh-B\'enard
convection in a simple fluid}
         {Phys. Rev. A}{46}{918-927}

\refarticle{refW102}
{\author{Jean-Marie}{Wersinger}, \author{John M.}{Finn}
  \and \author{Edward}{Ott}}{1980}
            {Bifurcations and strange behaviour in instability
saturation by nonlinear mode coupling}
         {Phys. Rev. Lett.}{44}{453--457}

\refartbook{refW103}
{\author{J.B.}{Walsh}}{1986}
            {An introduction to stochastic partial differential equations}
            {Ecole d'\'et\'e de probabilit\'es de St-Flour XIV}
            {\author{P.L.}{Hennequin}}
            {Springer}{Berlin}{pp266--439}

\refphd{refW104}
{\author{R.A.}{Weston}}{1989}
            {Lattice Field Theory and Statistical-Mechanical Models}
            {University of Cambridge}
 
\refarticle{refW105}
{\author{K.}{Wiesenfeld} \and \author{Bruce}{McNamara}}{1985}
            {Period-Doubling Systems as Small-Signal Amplifiers}
            {Phys. Rev. Lett.}{55}{13--15}

\refartbook{refW106}
{\author{K.}{Wiesenfeld}}{1989}
            {Period doubling bifurcations: what good are they?}
            {Noise in nonlinear dynamical systems, Vol. II}
            {\author{Frank}{Moss} \and \author{P.V.E.}{McClintock}}
            {Cambridge University Press}{Cambridge}{pp145--178}

\refarticle{refW107}
{\author{K.}{Wiesenfeld}}{1985}
            {Virtual Hopf phenomenon: A new precursor of
period-doubling bifurcations}
            {Phys. Rev. A}{32}{1744--1752}

\refarticle{refW108}
{\author{K.}{Wiesenfeld} \and \author{N.F.}{Pederson}}{1987}
            {Amplitude calculation near a period-doubling bifurcation:
        An example}
            {Phys. Rev. A}{36}{1440--1444}
\refarticle{refW109}
{\author{C.}{DeWitt-Morette} \and \author{K.D.}{Elworthy (eds)}}{1987}
            {New stochastic methods in physics}
            {Phys. Rep.}{77}{121--382}

\refbook{refW110}
{\author{L.C.G.}{Rogers} \and \author{David}{Williams}}{1987}
            {Diffusions, Markov Processes and Martingales.
        Vol 2: \ito\ calculus}
            {Wiley}{Chichester}

\refarticle{refY100}
{\author{Marc}{Yor}}{1989}
            {On stochastic areas and averages of planar Brownian motion}
            {J. Phys. A}{22}{3049--3057}

\refarticle{refZ100}
{\author{H.}{Zeghlache}, \author{Paul}{Mandel} \and \author{C.}{van den Broeck}
}{1989}
           {Influence of noise on delayed bifurcations}
           {Phys. Rev. A}{40}{286--294}

\refartbook{refZ101}
{\author{Paul}{Mandel}, \author{H.}{Zeghlache} \and
\author{T.}{Erneux}}{1989}
           {Time-dependent phase transitions}
        {Far from Equilibrium phase Transitions}
{\author{Luis}{Garrido}}{Springer}{Berlin}{pp218--236}

\endgroup
\inrefsfalse

\spaceandathird


\damtpaddress

\windowcover

\def\theauthor{G.D.~Lythe}

\title{Noise and dynamic transitions}

\signature

\abstract{ A parabolic stochastic PDE is studied analytically and
numerically, when a bifurcation parameter is slowly increased through
its critical value. The aim is to understand the effect of noise on
delayed bifurcations in systems with spatial degrees of freedom.
Realisations of the nonautonomous stochastic PDE remain near the
unstable configuration for a long time after the bifurcation parameter
passes through its critical value, then jump to a new
configuration. The effect of the nonlinearity is to freeze in the
spatial structure formed from the noise near the critical value.  }

\publishedin{pp 181-188 in 
{\sl Stochastic PDEs}, A. Etheridge (ed)\filbreak\noindent
 (Cambridge University Press 1994)}

\begingroup
\ShowTP
\endgroup

\pageno=1

\beginsect Introduction

Many physical systems undergo a transition from a spatially uniform
state to one of lower symmetry. Such systems are commonly modelled
by a simple differential equation which has a bifurcation parameter
with a critical value at which there is an exchange of stability
between steady states\addref{refN102}{N102}.
 The model can be improved by  adding noise\addref{refM103}{M103},
 which can be understood as allowing for small rapidly varying effects
neglected in the formulation of the model. Noise is also necessary to
provide the initial symmetry-breaking which permits the system to choose one of
the available lower-symmetry states.

If the bifurcation parameter is not a function of time, the effect of
noise is to make a bifurcation ill-defined in an $\Or{\ep}$
 region near the critical value, where $\ep$ is the size of the
noise\addref{refM113}{M113}.
 A more dramatic effect is found if the bifurcation parameter
is a function of time, corresponding to an experimental situation in
which the critical value is not known in advance,
so the parameter is slowly changed until a qualitative change is seen in the
behaviour of the system\addref{refM107}{M107}.

In the case of the pitchfork bifurcation, described by the normal form
$$\dot y = gy-y^3,
\eqno\myeqno\myeqlabel{pitchfork}$$
it is normally said that, if $g>0$, $y$ will be found either at 
$y=\sqrt{g}$ or $y=-\sqrt{g}$. However, on solving the stochastic
differential equation
$$dy = (gy-y^3)dt + \ep dw
\eqno\myeqno\myeqlabel{pitchforkwn}$$
with $g=\mu t$, starting with $y=0$ at $g=g_0<0$, it is found that
$y$ remains near $y=0$ until well after $g=0$ (Figure 1). 
(The Wiener process -- standard Brownian motion --
is denoted by lower case $w$.)
If $\mu$ is small, a typical
trajectory consists of a long sojourn near $y=0$, a sudden jump away
from $y=0$ and then relaxation towards $y^2=g$.
The value of $g$ at which the jump occurs is a random variable whose
probability distribution can be found by solving the linearised
version of \pitchforkwn\addref{refL100}{L100}; the mean value is approximately 
$\sqrt{2\mle}$ and the width of the \pd\ is proportional to $\mu$.

\figure 4truein 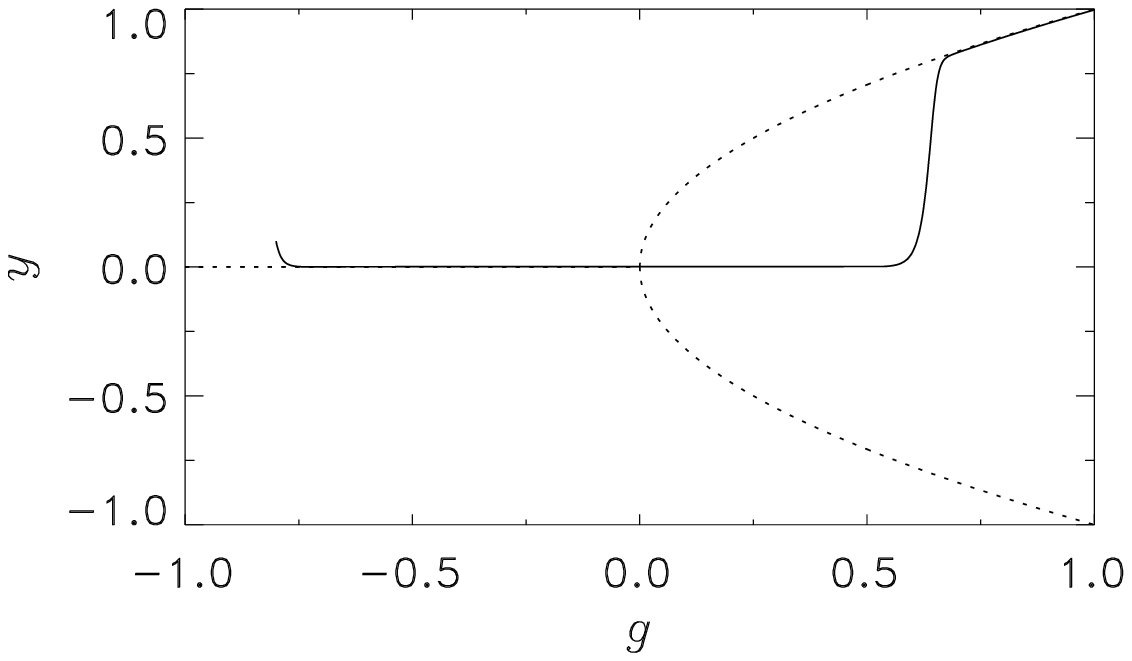
{\tenit Dynamic pitchfork bifurcation with noise.} The solid line is
one trajectory of the ordinary SDE \pitchforkwn, with $\ep=10^{-10}$,
$\mu=0.01$, and initial condition $y=0.1$ at $g=-0.8$.
Also shown, as dotted lines, are the loci of stable  fixed points
of the corresponding autonomous system \pitchfork.
The trajectory lingers near $y=0$ for a long time after $g$ passes
through $0$; the value of $g$ at which the jump towards $y=\pm \sqrt{g}$
occurs is a random variable with mean approximately 
equal to $\sqrt{2\mle}$.
\myfigurelabel{npf}

The phenomenon of delayed bifurcation just described 
and its sensitivity to noise has been studied
theoretically and experimentally in the case of 
non-autonomous stochastic ordinary differential
equations\addfourrefs{refS102}{S102}{refS109}{S109}
{refL106}{L106}{refA105}{A105};
the corresponding phenomenon, delayed transition, for PDEs is the
subject of this article. A dynamic transition is described by the following
 non-autonomous parabolic stochastic partial differential equation:
$$dY = (g(t)Y-Y^3+\Delta Y)dt + \ep dW,
\eqno\myeqno\myeqlabel{slspde}$$
where $g=\mu t$ is slowly increased through $0$,
 $W$ is the Brownian sheet,
$\Delta=\sum_{i=1}^m{\partial^2\over\partial x_i^2}$ is the Laplacian
in $\real^m$
(the derivatives exist only in the sense of generalised functions)
and $\mu$ and $\ep$ are constants such that
$$\ep\ll\mu\ll 1.
\eqno\myeqno\myeqlabel{mullep}$$

In section 2  this initial value problem is solved
 using periodic boundary conditions
in one space dimension, in which case $Y$ is a stochastic process with
values in the space of real-valued continuous functions on $[0,L]$,
and has continuous mean $\mean{Y_t(x)}$ and correlation function
$\mean{Y_t(x)Y_t(x')}$\addtworefs{refF103}{F103}{refG104}{G104}. 
In section 3 the corresponding equation in
 space dimension, $m$, greater than one is discussed.

The present work was motivated by the behaviour of some  nonlinear
systems of ordinary
and partial differential equations whose dynamics are controlled by
noise\addtworefs{refH103}{H103}{refM104}{M104}.
 The controlling influence of noise arises because trajectories
spend long times near a slow invariant manifold, and the dynamics near
this manifold are similar to those of a dynamic
bifurcation\addref{refL100}{L100}.
 The most important parameter in the description of these
systems is $\mle$, where ${1\over\mu}$ is the timescale for the
dynamics near the slow manifold and $\ep$ is the magnitude of the noise.

\figure 7truein  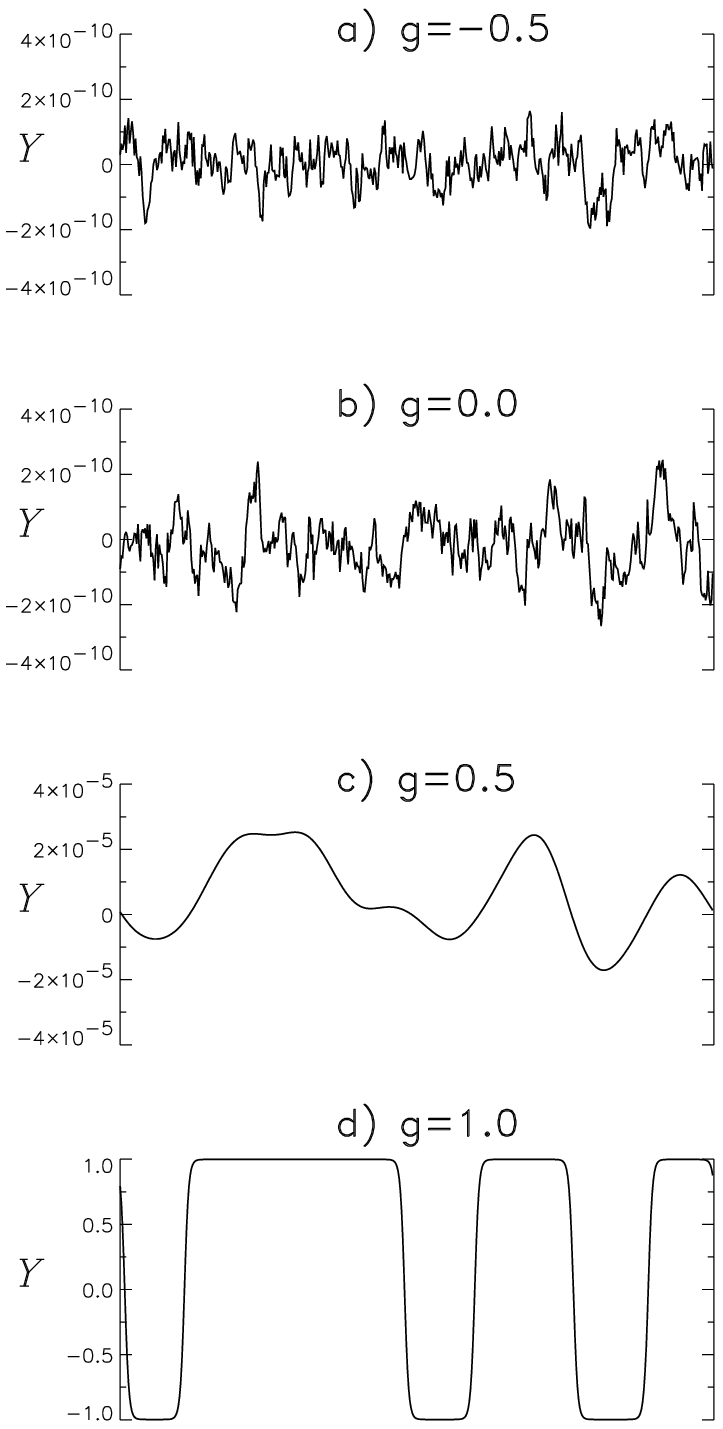
{\tenit Dynamic transition in one space dimension.} 
One numerically-generated realisation 
of the stochastic PDE \slspde\
$Y_t(x)$, $x\in[0,L]$, is displayed at four times as the
bifurcation parameter, $g=\mu t$, is
slowly increased, passing through $0$ at $t=0$. 
(Note the different vertical scales.) It remains 
close to the zero configuration until well after $g=0$. Nonlinear
terms become important when $g\simeq\sqrt{2\mle}$; their effect is
to freeze in the spatial structure formed, near $g=0$, from the
noise. The data shown here were generated using a finite difference
method with 500 points. ($L=300$, $\mu=0.01$, $\ep=10^{-10}$.)
\myfigurelabel{fcf}

\beginsect Model dynamic transition in one space dimension

Four configurations from a numerical realisation of \slspde\ with $m=1$
are  shown in Figure 2. 
According to static stability analysis
 the zero configuration is unstable for $g>0$, but $Y$ typically
remains much closer to $0$ than to $Y^2=g$ until well after $g=0$
when $g$ is time-dependent.
The first part of a dynamic transition, 
the loss of stability of the zero configuration, is thus
controlled by the noise and the sweep rate, not by the nonlinearity.
 
If $Y$ obeys \slspde,
the linearised version of the stochastic differential equation for its
$k$th Fourier mode, $u_t^{(k)} = {1\over \sqrt{L}}
\int_0^L Y_t(x'){\rm e}^{{\rm i}kx2\pi/L}dx'$, is
$$du^{(k)} = (g(t)-({2\pi\over L})^2k^2)dt + \ep dw^{(k)},
\eqno\myeqno\myeqlabel{zerothfmsde}$$
with each $w^{(k)}$ an independent Wiener process.
Thus, for as long as nonlinear terms are unimportant, each Fourier
mode evolves independently, satisfying an equation of the dynamic
bifurcation type.

As the in ODE case, the new configuration (Figure 2(d)) appears
abruptly at a value of $g$ which is a random
variable with mean approximately $\sqrt{2\mle}$.
The effect of the cubic nonlinearity
is to arrest the exponential growth of $\mean{Y_t(x)Y_t(x)}$
when it becomes $\Or{1}$ and to 
freeze in the spatial structure formed during the slow sweep.
Establishing when this happens makes it possible to answer another
question which is not considered in the static version:
what is the  typical size of the spatial domains formed in the transition?

To examine the dynamics of the loss of stability of the zero configuration,
consider the solution of the linearised version of \slspde,
$$dY = (g(t)Y+\Delta Y)dt + \ep dW,
\eqno\myeqno\myeqlabel{linslspde}$$
with initial data $Y_t(x) = f(x)$ on $x\in [0,L]$ 
at $ t=t_0<0$ when $g=g_0$:
$$Y(x,t) = \int_0^LG(t,t_0,x,x')f(x')dx' +
\ep\int_{t_0}^t\int_{[0,L]}G(t,t_0,x,x')dW_{t}(x').
\eqno\myeqno\myeqlabel{slspdesoln}$$
The first integral is an ordinary integral; the second a space-time
\ito\ integral. The fundamental solution
$G(t,t_0,x,x')$ is given by
$$G(t,t',x,x') = {1 \over \sqrt{4\pi(t-t')}}{\rm e}^{\ha\mu(t^2-t'^2)}
\sum_{j=-\infty}^{\infty}{\rm e}^{-{(x-x'-jL)^2\over4(t-t')}}.
\eqno\myeqno\myeqlabel{bigg}$$

The solution \slspdesoln\
prescribes the mean and correlation function of $Y$ as a
function of time. The 
mean value of $Y$ is the first term, an ordinary integral which, for
large $t-t_0$, is given by
$$\mean{Y_t(x)}={\rm e}^{\ha\mu(t^2-t_0^2)}\bar y
\eqno\myeqno\myeqlabel{meany}$$
where $\bar y$ is a smoothed version of the initial data $f(x)$.
In the rest of this article, it is assumed that $\bar y\le\Or{1}$ and
$2\mle< g_0^2$. This ensures that
$\mean{Y_t}^2\ll\mean{Y_t^2}$, ie 
$Y$ is taken as a mean-zero Gaussian process.
It is also assumed that $|t_0|>{1\over\sqrt{\mu}}$, so that quasi-static
equilibrium is reached before $g=0$, and $L\gg \sqrt{t-t_0}$ so that,
at any one $x$, only one term in \bigg\ is important.

The correlation function is a mean square quantity which 
can be evaluated using the `delta function' property of 
the Brownian sheet as an integrator:
$$\eqalign{
\mean{Y_t(x)Y_t(x')}
&=\ep^2\int_0^L\int_{t_0}^tG^2(t,s,x,x')dsdx'\cr
&= \ep^2
\int_{t_0}^t ds {{\rm e}^{\mu(t^2-s^2)} \over \sqrt{8\pi(t-s)}}
 {\rm e}^{-{\,\,\,(x-x')^2\over8(t-s)}}.}
\eqno\myeqno\myeqlabel{corr}$$

For  times $t$ such that $-t>1/\sqrt{\mu}$ the 
nonautonomous correlation function \corr\
 differs only by $\Or{\mu}$ from the
result obtained for fixed $g<0$ if $t_0\to\infty$\addref{refH107}{H107}:
$$\mean{Y_t(x)Y_t(x')} = {\ep^2 \over 4\sqrt{|g|}}
{\rm e}^{-{{|x-x'|\sqrt{|g|}}}}.
\eqno\myeqno\myeqlabel{qs}$$
In the non-autonomous case, the correlation function
 remains finite as $t$ passes through $0$, and
 for large positive times ($t> 1/\sqrt{\mu}$), it is well approximated by:
$$\mean{Y_t(x)Y_t(x')} \simeq
\ep^2{{\rm e}^{\mu t^2} \over \sqrt{8\mu t}}{\rm e}^{-(x-x')^2/8t}.
\eqno\myeqno\myeqlabel{eto}$$
Note the Gaussian form of the spatial correlation and the
existence of a
 characteristic length $\sqrt{8t}$ at time $t$.
This formula is valid until $g\simeq\sqrt{2\mle}$, when the cubic nonlinearity
becomes important. 
The correlation length at this time becomes the
characteristic size of the spatial domains formed (Figure 2(d)).

\beginsect Higher space dimensions

Proceeding as in section 2, the expression obtained for
the correlation function at time $t$ is
$$\mean{Y_t(x)Y_t(x')}=
\int_{t_0}^t ds {{\rm e}^{\mu(t^2-s^2)} \over (8\pi(t-s))^{m\over 2}}
 {\rm e}^{-{||x-x'||^2\over 8(t-s)}},
\eqno\myeqno\myeqlabel{dgocorr}$$
where $||x-x'||$ is the Euclidean norm on $\real^m$.
The major difference from the case $m=1$ is that
 the correlation function does not approach a finite
limit as $||x-x'||\to 0$. This is a general feature
of second order stochastic PDEs in more than one space
dimension\addtworefs{refW103}{W103}{refD104}{D104}.
On a finite grid, equation \linslspde, with $g<0$ fixed,
is of interest because it generates a stochastic process with non-zero
correlation length in both space and time\addref{refG108}{G108}.
The singularity in the correlation function at $x=x'$
manifests itself in the normalisation,
 which increases without
 a finite limit as the grid spacing is decreased.
In the nonautonomous version the same divergence is seen
(logarithmic when $m=2$), but
for $g>0$ the singularity is only
noticeable on very small length scales. The nature of these divergences,
in both discrete and continuous versions, is conveniently studied
in Fourier space, where the linearised equation 
separates into uncoupled ordinary SDEs, each of which can be solved exactly.

The finite difference method for a parabolic SPDE consists of
replacing the infinite dimensional system \slspde\ by $N^m$
ordinary SDEs on a grid of equally-spaced points in $[0,L]^m$.
For the SPDE \slspde, the SDE at each point is
$$dY(x) = (gY(x)-Y^3(x))dt + \rho\tilde\Delta Y(x)dt
+\rho^{m\over 4}\ep dw
\eqno\myeqno\myeqlabel{fdiff}$$
where the discrete Laplacian $\tilde \Delta$ is defined by
$$\tilde\Delta Y(x) = \sum_{x'}Y(x')-2mY(x)
\eqno\myeqno\myeqlabel{dislap}$$
and the sum is over the $2m$ nearest neighbours of $x$.
Each of the $N^m$ ordinary SDEs has an associated Wiener process,
 independent  of all the others, and so can be numerically solved
in the normal way\addref{refK100}{K100}.
 Note the scaling of the magnitude of
the noise added at each grid point with $\rho$.
The smallest length scale resolved is roughly
the distance between nearest neighbours, ${1\over\sqrt{\rho}}$.
As in numerical solution of
 deterministic parabolic PDEs, there is
 a maximum timestep, proportional to ${1\over\rho}$, if the 
finite difference method is to be numerically
stable\addref{refL113}{L113}.

Solved on a finite grid in this way, the behaviour of realisations of
the non-autonomous version SPDE \slspde, with $m>1$ and 
$g=\mu t$ as in section
2, is qualitatively similar to that in one space 
dimension\addref{refL116}{L116}.  That is,
a Gaussian form of the spatial correlation function emerges from the
slow sweep through $g=0$,
$$\mean{Y_t(x)Y_t(x')} \simeq
\ep^2\sqrt{{\pi\over\mu}}
{{\rm e}^{\mu t^2} \over (8\pi t)^{{m\over2}}}{\rm e}^{-||x-x'||^2/8t},
\eqno\myeqno\myeqlabel{etogenm}$$
and metastable domains of positive and negative $Y$ are formed when
the nonlinearity becomes important, at $g\simeq\sqrt{2\mle}$.

\beginsect Conclusion

In systems with spatial degrees of freedom, the phenomenon 
of delayed bifurcation is accompanied by a noise-controlled formation
of spatial structure. 
Here the stochastic PDE \slspde\ was studied, whose realisations
consist long sojourns near
the zero configuration during which the shape of the 
spatial correlation function becomes Gaussian,
 followed by an abrupt change to 
the nonlinear \regime, with the formation of
domains of positive and negative $Y$.
This spatial structure, formed near the critical value of the
bifurcation parameter $g$
when noise dominates, is frozen in at
$g\simeq\sqrt{2\mle}$, when the characteristic length
is $\sqrt{8t}$.

%

\showrefs

\Trailers

\end